\documentclass[12pt,a4paper]{article}
\usepackage{amssymb}
\usepackage[T2A]{fontenc}
\usepackage[cp1251]{inputenc}
\usepackage[russian,english]{babel}
\usepackage{graphicx}
\usepackage[dvips]{psfrag}
\usepackage[unicode]{hyperref}
\begin{document}
\begin{center}
{\Large 
A FAREWELL TO PARTICLES}\\[3mm]
{B.~P.~Kosyakov}\\[3mm]
{{\small Russian Federal Nuclear Center--VNIIEF, Sarov, 607188 Nizhni{\u\i} Novgorod Region, Russia\\
}} 
\end{center}
\begin{abstract}
{We outline the course of affairs in the experimental and theoretical fields of nuclear and particle physics which determined its finale, and give several fragmentary remarks on its present state.
The essay tells about events and their participants, known from the literature, but presented here  from the perspective of a person whose 50-year labor activity, 1972--2022, proceeded at the All-Union 
Scientific Research Institute of Experimental Physics, Sarov.
In order to predict the fate of particle physics and related astrophysics and cosmology, it is useful to become aware of the facts about another branch of physics that has already gone through its decline $\sim 30$ years ago, the physics of nuclear weapons.
These facts are important not in themselves, but as evidence of the growing problems of science and social life, which are not only far from a satisfactory solution, but have not even been the subject 
of any serious discussion.
}
\end{abstract}

\section{What is this paper about?}
\label
{Introduction}
The title of the paper is a paraphrase of the famous novel {\it A Farewell to Arms} by Ernest Hemingway, in turn, borrowed from the title of a sonnet by the 16th century English playwright George Peele.  
The era of research related to elementary particle physics and its theoretical foundation, quantum field theory, is nearing completion.
The proscenium is captured by astrophysics and cosmology.
The careful outside observer can judge the change in the mainstream  by the achievements that have been awarded the Nobel Prize over the past couple of decades.
This fact, of course, is not proof, but is quite suitable as a sign.
A more compelling argument is as follows.
Physics, with due reservations, is an experimental science.
Great experimental discoveries shock us with their unexpectedness.
From the Galilean experiments, we learned that the free fall of any bodies occurs in a uniform way, regardless of their mass; Rutherford's experiments taught us that the atom consists of a massive nucleus $\sim 10^{-13}$ cm in size, around which light electrons revolve at distances of $\sim 10^{-8}$  cm, in other words, the substance consists mainly of voids; and from relatively recent astronomical observations, we found that the Universe is expanding with increasing acceleration.
A science in which the flow of experimental surprises dries up leaves the spotlight.
This is exactly what is happening with  elementary particle physics.  

I will attempt, to the best of my knowledge and understanding, to outline the course of affairs in the experimental and theoretical fields of particle physics, which determined its finale, and I will give a few fragmentary remarks concerning its present state.
The essay tells about the events and their participants, known from the literature, but presented here from the perspective of a person whose 50-year labor activity, 1972--2022, proceeded at the All-Union Scientific Research Institute of Experimental Physics, Sarov~\footnote{The original name of the organization is Design Bureau No 11 (KB-11).
It was located in the village of Sarova, Mordovian Autonomous Soviet Socialist Republic, which became a closed facility.
Now it is the All-Russian Scientific Research Institute of Experimental Physics (VNIIEF).
The settlement grew into a city, at one time bearing the name Arzamas-16. 
Now it is Sarov.
The city still remains closed.}.
For the first 20 years, I was involved in the development and improvement of nuclear weapons, and therefore could devote only a small part of my time and energy to studying current problems in field theory and particle physics. 
However, since 1991, I moved away from the defence affairs and completely immersed myself in these problems.
I was familiar with a number of Russian and foreign scientists, the heroes of the epic under discussion, and I present information received first hand.
Otherwise, I give links to its sources.

Scientific researches are carried out in specific historical conditions.
Their results and consequences are determined not only by the objective logic of the development of the subject, but also by a variety of subjective circumstances.
In order to predict the fate of particle physics and related astrophysics and cosmology, it is useful to become aware of the facts about another branch of physics that already went through the decline $\sim 30$ years ago, the physics of nuclear weapons.
Having not been a bearer of state secrets for more than a quarter of a century, I can speak openly about a number of things discussed both in scientific literature and on the Internet. 

The breadth of topics covered hopefully will not scare the reader away.
If one does not agree with the point of view expressed here on any specific issues, then I am ready to listen carefully to criticism -- albeit harsh, but clear and reasoned.

\section{Breakup milestones}
\label
{History}
The US Congress decided to stop building the Superconducting Super Collider (SSC) on October 19, 1993.
By this time, more than 2 billion dollars had been spent, 23 km of the tunnel and 17 shafts from the surface had been dug, the construction of buildings and underground experimental halls has begun; in short,  $\sim 20 \%$ of the total amount of work has been completed. 
The SSC project was to construct a proton-proton collider 87 km in circumference. 
It was planned to produce a collision energy of $20\times20$ TeV and beam luminosity of $10^{33}$ cm${}^{-2}$s${}^{-1}$. 

The official reason for the closure of the project was a large overestimation, jumped from the initial cost estimate of $\$ 5.9$ bn to the requested at the end of $\$ 11$ bn.
Meanwhile the decision was made some two weeks after the shooting of the House of Soviets in Moscow, October 3--4, 1993.
So the real reason, in my opinion, is quite clear: the science developed in the USSR ceased to exist.

The race in the study of elementary particles, to put it bluntly, is a tribute to the fashion of the 20th century.
By coincidence, the scientific community rushed into the depths of the microworld, not, say, biology. 
Should America continue to spend tens of billions of dollars pretending to have a selfless passion for understanding particle physics when there is no more rival?
To whom to brag about the Nobel Prizes?
Jesus taught: ``Do not cast your pearls before swine''.

This attitude is leading to the gradual closure of large accelerator system in the United States.
The Stanford Linear Accelerator (SLAC), the world's largest 50 GeV electron accelerator, is being repurposed as an X-ray source for applications in solid-state physics, chemistry, biology and medicine.
Тhe Tevatron, a proton-antiproton collider with a total center-of-mass energy of 1.96 TeV, completed  operation in 2011.
The only large facility, the Relativistic Heavy Ion Collider (RHIC), designed to accelerate gold ions to 100 GeV per nucleon to form a quark-gluon plasma in ion collisions, operating from 2000 to the present, is on the verge of shutdown.

In 2008, CERN launched the Large Hadron Collider (LHC), a ring colliding particle accelerator with ring circumference of 26.7 km, designed to accelerate protons to 6.5 TeV or lead ions to 1.4 TeV per nucleon.
The European LHC project was implemented 15 years after the cancellation of the American SSC  project, but has a more modest performance.
Should we be surprised here?
As Mikhail Yur'evich Lermontov once remarked, ``fashion, starting from the upper strata of society, descended to the lower ones, who wear it out''.

Scientists convinced European governments to give them 6 billion euros to build the LHC to experimentally test the existence of the ultimate building block of the Standard Model of particle physics, the scalar Higgs boson, and find out whether the head-on collision of heavy ions actually turns ``normal nuclear matter'' into quark-gluon plasma, or their American colleagues on RHIC wishful thinking.
Both tasks were successfully completed, and physicists began to venture colliders with higher energies and luminosities. 
After the LHC, they pursued to build the Future Circular Collider (FCC), 80--100 km in circumference,  to accelerate colliding proton beams to an energy of 100 ТeV,  construct an electron-positron factory with a beam energy of 45--175 GeV, and their combination, opening the way to studying all kinds of hadronic, lepton and hadron-lepton reactions \cite{CERN}.

Are these plans viable? 
Imagine the following semi-fantastic plot.
An authoritative physicist (P), a major European official (O), and a wealthy Maecenas (M) meet. The atmosphere of the meeting is relaxed, conductive to frankness.
P is trying to figure out how to fund the FCC project.
The conversation begins:

\noindent
(M): What is the expected scale of costs for the construction of the FCC?   

\noindent
(P): The estimate I hope will not exceed 10 billion euros.
The financial burden can be reasonably distributed across countries and companies, and stretched out over time, if we start preparing today.
The Russians will certainly contribute, definitely the Chinese, most likely the Japanese, quite possibly the Americans. 
It will fall from everywhere~-- from Australia, Canada,  India, Korea.

\noindent
(M): The Russians? 

\noindent
(P): 
Yes. 
As far as I know, Russia's participation in the construction of the 
International Thermonuclear Experimental Reactor in France continues.
Doesn't the FCC worse?

\noindent
(O): What do you intend to discover with such an expensive tool?  

\noindent
(P):  We will attempt going beyond the Standard Model of elementary particles. 

\noindent
(M):  Why?
Doesn't the very word ``standard''  a synonym for the word ``perfect''?

\noindent
(P): No, the model is far from perfect.
It is not a theory of the unity of the three fundamental interactions of nature: strong, electromagnetic, and weak.
Take, for example,  the fact that the internal symmetry group of the Model, ${\rm SU}(2)\times{\rm U(1)}\times{\rm SU}_c(3)$, is not semisimple, and hence there is not a single gauge coupling constant...   

\noindent
(O): Wow, the Model has bad karma! 
Will the FCC bring these interactions together?  

\noindent
(P): Of course not.
The energy of the particles accelerated by the FCC is far from enough to get to the merging of these three interactions. 
The merger is expected somewhere around  $10^{14}-10^{16}$ GeV.
However, the paths of the three running interaction constants will intersect at one point in this region if there is supersymmetry in the microcosm \cite{Wilczek}, in other words, if all the observed particles have heavy superpartners.  
Unfortunately, no traces of supersymmetry have been found at the LHC...

\noindent
(O): And how do you order clearly and convincingly explain to the average 
European that 10 billion euros from his taxes need to be paid for the making of the FCC just because the priests of pure reason fail to reduce their chakras to the point of $E=10^{16}$ GeV? 

\noindent
(P): Humor appreciated! 
However, let's switch to a more businesslike tone.   
 
\noindent
(O): Well, let's go.
Everything is known in comparison.
Otto Hahn and Fritz Strassmann found that an ${}^{235}{\rm U}$ nucleus can be split by a neutron into two fragments, plus 2 or 3 neutrons, with releasing 200 MeV of energy.
Thus, a source was discovered whose energy is $10^8$ times greater than the energy of chemical sources based on breaking molecular or atomic bonds.
Soon, nuclear bombs and nuclear power plants came into existence.
The appearance of these things is the second most significant event in history if the exclusion of Adam and Eve from Paradise is taken to be the first.
Then a person lost personal immortality but retained the hope that life will continue in descendants. 
Now the price of knowledge is even more expensive. 
Careless {homo sapiens} can disappear from the planet altogether. 
To live in a new reality, mankind needs to grow up quickly.
So, the applied side of nuclear physics seems clear to everyone!
In the late 1940s, a boom in experimental discoveries of elementary particles began.
To date, a great many of them have been discovered, and their properties have been studied in detail.
What are the practical fruits of research? 
How did knowledge about particles affect our everyday life and technology? 

\noindent
(P): Leaving aside the cultural and scientific world-view significance of the discoveries...

\noindent
(O): Please give this honor to the humanities; let's talk about the utilitarian topics.

\noindent
(P): OK. I want to draw your attention to two important things.
First, we have come to understand that leptons have no internal structure; today they considered elementary.
In contrast, hadrons are composite systems.
They are built from quarks.
There is a strong belief \footnote{Although rigorous, generally accepted, evidence for this concept is missing.} that it is impossible to extract a single quark from a hadron; in an isolated form, a quark is not observed.
In any case, there is no need to pin hopes on the discovery of new sources of energy associated with the splitting of hadrons.
The most capacious supplier of energy is the annihilation of particles and antiparticles.
However, this source is practically inaccessible to us since antimatter is absent in natural conditions (baryon asymmetry reigns in our world), and the artificial accumulation of antimatter requires huge energy costs.
Secondly, knowledge of the laws of the microworld removes fears about real and imaginary dangers lurking in microworld.
For example, before testing the first hydrogen bomb on Eniwetok Atoll, a close look has been given at the probability of initiating a sustained thermonuclear detonation of deuterium dissolved in ocean water.
Before the launch of the RHIC, there was a debate in the media and scientific literature as to whether a quark-gluon plasma could turn into a strangelet~\footnote{Strangelet is a hypothetical object representing a bound state of $u$, $d$, and $s$ quarks in equal shares~\cite{Bodmen, Witten}.
Such objects are elements of strange matter.
It is assumed that it is more stable than ordinary matter, and hence the contact of a strangelet with ordinary matter renders it strange matter.}, which would threaten the Earth to become its strange counterpart~\cite{Dar, Jaffe}.
Prior to the launch of the LHC, evidence was presented to the general public that if a microscopic black hole will be born in a collision of protons, the  black hole will not suck the entire Earth, or at least Lake of Geneva.
                
\noindent
(O): However, back to colliders.
You do not raise the question of a direct experimental study of physics in the Grand Unification region $E\sim 10^{15}$ GeV.
This is not surprising.
With the current level of development of accelerator technology, in a section 1 km long, the maximum proton acceleration reaches $\le 10^3$ GeV.
Thus the circumference of a collider designed for collisions at $10^{15}$ GeV should not be less than 1 light year, {that is}, $10^{13}$ km.
You are trying to capture the echoes of processes from the area of the Great Unification by registering the rarest deviations in ordinary processes occurring at energies of the order of 1--100 TeV.
The probability of a rare phenomenon can be estimated from the uncertainty relation for energy fluctuations $\Delta E$ and their implementation time $\Delta t$, which (in natural units $\hbar=1$, $c=1$) has the form $\Delta E\cdot\Delta t\sim 1$.
With $\Delta E=10^{15}$ GeV, we find that such an extremely transient phenomenon occurs during $\Delta t\sim 10^{-40}$ c.
In other words, the probability that it will happen within a given second is $10^{-40}$.
To increase the statistics, you need to repeat the situation with a large number of participating particles.
Translated into everyday language, this means that the processes observed on modern colliders will tell us something interesting about the Grand Unification, if we are ready to burn a tremendous amount of electricity~\footnote{For reference:  the LHC in operation consumes $10\%$ electricity for the entire canton of Geneva.}.
Will the European ``consumer society'' make such a sacrifice on the altar of knowledge?

\noindent
(M):  Look here! If the collider has become empty fun, why have the pragmatic Americans not closed the RHIC for 22 years, although in papers on the study of quark-gluon plasma one cannot catch faint hints of discoveries that could qualify for the Nobel Prize?
Let me remind you that the operation of the RHIC annually costs 100-150 million dollars; and the total costs likely exceeded $\$3$ billion.

\noindent
(P): 
This is all the more strange considering that the same studies at the LHC are carried out at higher collision energies, using ${}^{82}{\rm Pb}$ nuclei, which are heavier than ${}^{79} {\rm Au}$.
It is reasonable assume that someone (not the military?) cherishes the hope of discovering a new mechanism for obtaining energy from the phase transition of nuclear matter into a quark-gluon plasma or from the decay of quark-gluon plasma into numerous relativistic hadrons.
Indeed, the temperature of these transitions is $\sim 200$ MeV, which is $10^4$ times higher than the temperature at the explosion of a thermonuclear bomb.
It is possible that the advance on the energy scale by 4 orders of magnitude serves as an incentive to continue funding this purely scientific research.

\noindent
(O): 
With a free understanding of your words, you do not seem to exclude the possibility that a heavy nucleus can not only be divided into two fragments, plus 2--3 neutrons, but even completely split into its constituent nucleons if this nucleus is turned into a lump of quark-gluon plasma.
However, after all, the initial stage requires the expenditure of energy for the acceleration of colliding nuclei, which is much higher than the energy of the decay products of quark-gluon plasma.
Where is the energy benefit?

\noindent
(P): 
The informal nature of our conversation allows for loose analogies between nuclear physics and some as yet vague phenomena in quantum chromodynamics.
Nuclear fission was inconceivable until the discovery of a ``projectile'' that easily penetrates the nucleus and violates its stability, the neutron.
A projectile of this kind, capable of entering a nucleus, excite it enough to turn it into a lump of quark-gluon plasma is not yet known.
We can make a guess as to whether a glueball, a particle that does not contain quarks but consists of only gluons, is suitable for this role.
The existence of this particle, predicted at the dawn of the development of quantum chromodynamics, has not been experimentally established with certainty.
Glueballs are electrically neutral, colorless, and have zero weak coupling constants.
It is  a bound hadronic state immune from the interaction with the environment; it is capable of unobstructed penetration to a nucleus.
To enter into a strong interaction, the glueball must split into two gluons, {that is}, undergo deconfinement.
However,  the confinement-deconfinement phase transition means a rearrangement of the vacuum structure, and this, in its turn, can render the nucleus, inside which the glueball is stuck, a lump of quark-gluon plasma.

\noindent
(O): ``Projectile'', you said, is not discovered.
I'm not sure that this discovery bodes well for us.

\noindent
(M): One may get the impression that we have approached the limit of spending on things that, by their nature, are divorced from vital human needs.
This has happened more than once in history. 
For example, the construction of the pyramids in ancient Egypt reached the upper bound of the expenditure of resources in erecting the pyramid of Cheops.

\noindent
(O): 
Oh, yes!
The lessons of history are important to remember.
Modernity has given rise to many activities which are useful by design, but do not represent the categorical imperative, for example, playing sports.
The results of physical competition sooner or later reach natural limits, and then professional sports arise.
People sacrifice their health and cripple their own and other people's destinies.
However, here, too, achievements are quickly going to the limit.
And doping comes into play.
I am far from direct comparisons of sports and the science of the microworld.
But still...

Of course this is a metaphor.
In life, such a meeting is filled with decent silences, but that does not make its upshot any less cruel.

At the end of the 20th century, another branch of physics collapsed.
Very little has ever been written on this event~\footnote{The initial stage of the ``path to the pier'' is described in \cite{SakharovMem}.}.
However, it would be, I think, not out of place to mention it too.

\section{A farewell to arms}
\label
{Farewell_to_Arms}
The 50th session of the UN General Assembly concluded the Comprehensive Nuclear Test Ban Treaty~\footnote{For brevity, we will use the term ``Treaty''.} on September 10, 1996~\cite{resolution}.
The Treaty is of an indefinite nature and provides for withdrawal from it only in the event of a threat to the highest national interests.

How could this happen so that the governments of all the leading nuclear powers would simultaneously be filled with peacefulness?
This event, in my opinion, was triggered by two circumstances.

First, the search for new prospects for nuclear weapons in the United States and the USSR has attained saturation.
Further advancement in this science became a waste of energy.
All conceivable versions of the physical schemes of nuclear and thermonuclear devices were already examined in calculations and nuclear tests, and the reliability and safety of the produced systems have reached an acceptable level.

The critical point was the failure of Ronald Reagan's Strategic Defense Initiative, in public discussions called ``Star War Program''.
It became obvious not only to scientists but also to the most conservative politicians that  any country cannot develop a combat X-ray laser in near space in the foreseeable future~\footnote{The task defied technical solutions.
I witnessed those events, but I did not take part in the enterprise, and I did not have access to the pertinent documents.
In my opinion (and my assessment was shared by many of my colleagues, more experienced in the development of nuclear weapons), nothing could come of either the Americans or us.
Veterans of this story refrain from commenting in the open literature.}.

Secondly, the West was in euphoria from the success in pacifying the ``Evil Empire'' whose political leaders themselves destroyed the Warsaw Pact, and then the USSR.
An unilateral moratorium on nuclear testing was announced by the USSR in 1990.
There seemed to be no limit to the pliability of the Russians.
  
Before the conclusion of the Treaty, the UK and France defended the right of nuclear countries to periodic conducting nuclear tests to maintain the combat readiness of existing arsenals.
This initiative was rejected by other members of the nuclear club, as well as non-nuclear countries.
Whether this happened out of short-sightedness or out of opportunistic considerations, I don't know.
But the very proposal seems reasonable, and I hope we can still return to it.

Nuclear weapons, made half a century ago, is not reproducible because it is impossible to use the previous technologies.
Changed metallurgy, machining, spraying, welding, robotics, chemistry, and nano-engineering.
Many manufacturing processes are prohibited as harmful to the health of personnel or environmentally unacceptable.
A number of technological chains, debugged in the USSR, were destroyed in the 90s.
Although the United States were not subject to such economic shocks, it also experienced a massive restructuring of the industry associated with the end of the arms race and the reduction of government orders for nuclear products.

A nuclear device machined in Russia or the United States to replace a device with an expired regulatory period may not be equal to its predecessor in the required operational properties.
The operation of a nuclear device depends on a variety of nonlinear physical processes that are sensitive to initial conditions, in particular to the factory design of this device.
Some nuclear systems among those intended for replacing the systems on combat duty must be tested three times a century.
If this is not done the combat unit may well lose its function -- to be {real means of restraint}.

Another argument is that the creative activity of a scientific specialist falls  within the interval 30--45 years.
Passing the baton from a generation of specialists with experience in development, testing, and commissioning nuclear weapons systems to the new generation is urgent.
There are few veterans of the Soviet nuclear era who continue their labor activity.
The situation in the USA is the same.

The development of nuclear weapons is not a science in the strict sense of the word.
It contains an element of art.
Systematically take into account all the essential physical processes in the operation of a nuclear device is unfeasible even taking into account the great achievements in calculations on modern computers.
The profession can be taught only in practice, having gone all the way from the development of a physical scheme to nuclear tests.
The role of an experienced mentor can be critical here.

The Treaty was signed by 183 states.
Among the 44 nuclear-capable states, India, North Korea, and Pakistan did not sign.
The Treaty was ratified by 164 states.
Of the ``list of 44'' countries, China, Egypt, Israel, India, Iran, North Korea, Pakistan, and the United States have not ratified the Treaty.
The Treaty has not officially entered into force.
Hovever, all countries are currently complying with the waiver of full-scale nuclear tests.

Experience shows that the human kind is not ready for complete nuclear disarmament.
Thanks to the nuclear and missile balance, it was possible to prevent the outbreak of  a third world war for almost eight decades.

The Treaty is likely to be easier to sign and ratify by all countries if its text is revised taking into account the above-mentioned proposal of the UK and France.
Of course, the possibility to test any new or even slightly advanced nuclear weapons systems is out of the question.
We should only discuss the verification of the combat properties of some selected nuclear devices in service of nuclear countries in full-scale underground tests, and under proper international control.

Now is not the right time for such initiatives.
The confrontation between the West and Russia has taken on a dangerous character and is approaching its culmination.
Therefore, talking about the resumption of nuclear testing may look like a provocative attack by one of the rabid nuclear scientists.
However, I would ask the reader not to rush to conclusions.

A shock wave is always followed by a rarefaction wave.
The world will calm down.
And then, perhaps my arguments will be remembered and regarded as worthy of discussion.
I'm not sure I'll live until then.
However, this is not so important. 
I fulfilled my duty.

\subsection{What's then? What's then?~\cite{Evtushenko}}
\label
{What's then}
Lavrenti{\u\i} Pavlovich Beria laid the foundations of the nuclear weapons complex of the USSR, closely following the successfully functioning American model.
For example, the development of new types of nuclear weapons and improvement of the existing ones in VNIIEF parallel the designation of the Los Alamos National Laboratory (LANL)~\footnote{The KB-11 workers sometimes jokingly referred to their facility as Los Arzamas among themselves.}.
There are also many similarities in the structure and staffing of LANL and VNIIEF.
But there were also serious differences, for example, the amount of funding.
Later, another nuclear center set up in the United States, the Lawrence Livermore National Laboratory (LLNL), and the USSR opens the All-Union Research Institute of Instrument Engineering (VNIIP) based in Chelyabinsk-70~\footnote{Now it is the Zababakhin All-Russian Research Institute for Technical Physics (VNIITF).
Chelyabinsk-70 has been renamed to Snezhinsk.}.

Since the signing of the Treaty, the original functions of these institutions have been lost~\footnote{A ``pedestrian'' probably needs to be explained that it is impossible to make a new type of nuclear weapons without full-scale testing.
Common sense and some experience with mechanical devices suggest that it will not be possible to produce even a combat-ready modification of conventional small arms without recourse to series of tests.
What can be said about much more complex and ``capricious'' nuclear weapons systems?}.
Today there are three major nuclear weapons centers in the US: LANL, LLNL, and Sandia National Laboratories (SNL) with a total permanent staff of $\approx 25,000$.
According to Wikipedia, the total budget of these centers in 2015 exceeded $\$6$ billion.
Spending on this scale over 30 years~\footnote{The US stopped its nuclear tests in 1992.} needs some explanation to the taxpayer.

From the previous tasks, there remains the maintenance of the combat capability and safety of already produced nuclear weapons.
However, to solve these problems, it would seem that great institutions and billions of dollars are not needed.
A small group of narrow specialists who know the design of nuclear weapons is sufficient. 
On the other hand, if the Treaty is violated in the foreseeable future (and this is quite acceptable) then how to respond to such a challenge without having a team of weapons developers, in particular specialists with real experience in the development and testing of nuclear devices?

LANL, LLNL, and Sandia have been constantly supplied with the most advanced computing technology.
Finally, supercomputers from the TOP500 list~\cite{TOP500} were installed there to calculate the physical processes peculiar to the explosion of nuclear weapons in the framework of the Advanced Simulation and Computing Program.
Note that the nuclear and thermonuclear weapons of both the USSR and the USA were developed through the use of one-dimensional or two-dimensional calculation codes.
At that time, the power of computers was two orders of magnitude lower than that of today's smartphones.
To describe a thermonuclear explosion in three-dimensional calculations, a computer with a power of over 100 TFLOPS is needed.
In 2005, with the commissioning of the ASC Purple computer at LLNL, this program achieved its goal.

Thereafter, how to give proof to the ``theorem of existence'' of the LANL, LLNL, and SNL?
The only compelling argument here remains: ``Well, the arms race is over...
However, Russia keeps its nuclear centers!''

This argument was probably foreseen by Washington when it demanded in the 1990s that the Kremlin continue to feed Arzamas-16 and Chelyabinsk-70, because the Yeltsin government in regarding this Soviet legacy was reminiscent of the rooster from the fable who found the pearly seed and says: ``Where is it? What an empty thing!''~\cite{Krylov}.
The US remained the only superpower.
They could not allow nuclear technologies and scientists to spread across the continents~\footnote{They fully understood the seriousness of the situation in the Russian nuclear weapons complex after the suicide of Vladimir Zinovievich Nechay, director of VNIIP,  on October 30, 1996, stemming from the fact that he failed to pay off the Institute's debts.}.
Although the Americans were masters of the situation, they emphatically did not want to use this to pump out information about Soviet nuclear weapons or poach Russian nuclear specialists.
There were plenty of opportunities for this (It is sufficient, I think,  to recall the story of Bakatin's gift to American intelligence services).
The main task was not to let anyone into Russia's nuclear secrets altogether~-- not partners in the nuclear club, the British and French, nor potential opponents, the Chinese and Indians.
From the intelligence data, VNIIEF and VNIITF specialists knew everything fundamentally important about the achievements in LANL, LLNL, and SNL.
Our colleagues overseas were just as aware of our state of affairs~\footnote{Information about nuclear and thermonuclear weapons of the USA and the USSR has always been a system of communicating vessels.
At the initial stage, the Russians swallowed the threshold amount of information about the atomic bomb.
Then the American ``bomb-makers'' did not hesitate to ``crib the answer'' from the Soviet notebook.
The subsequent story was smoother and more balanced; a nuclear specialist might say to his antipode: ``I know that you know that I know''.}.
After the cessation of testing, the level of knowledge about physics of nuclear weapons remains frozen in both countries.
Therefore, Russian nuclear weapons secrets are of little interest to US specialists, in general this stuff is known to them.

The United States has donated $\sim \$ 100$ million worth of technologies and funds to Russia to make a system for monitoring fissile materials in Russian nuclear centers.
As for the problem of retaining nuclear scientists, it has been solved in a completely scientific way.
It is well known that statistical fluctuations in a system of $N$ particles are proportional to $\sqrt{N}$.
For example, in an organization with 10,000 employees, there are $\sim 100$ people responsible for the problem.
They are active and hence easily detected.
The rest of the masses live within the average flow; it is sufficient to feed them a little suggesting that everyone around is much worse.
The highlighted $\sqrt{N}$ group should be divided into two parts.
The first includes people of the power-hungry type.
They need to be seated in the chairs of the chiefs.
Then they will become not just manageable but will think without prompting what is required of them.
The second part consists of people who are active and creative but ``left-handed'', that is, not capable of a bureaucratic career.
For them, an International Science and Technology Center (ISTC) should be founded, the official goal of which will be ``providing opportunities for the weapons scientists in Russia to redirect their talents to peaceful activities, and promoting their integration into the international scientific community''.
They themselves will come there with their ideas and take part in the competition of projects.
Financing their projects (modest equipment, international travels, grant payments $\le \$ 300$ per month), will lead the restless mind from dangerous directions, at least the Russian footcloth will not be rewound into a Muslim turban.
In 1992, the United States initiated the opening of such a center in Moscow.

Russian President Dmitri{\u\i} Anatolyevich Medvedev stopped Russia's participation in it in 2010 without any explanation.
Before that, all other foreign charitable foundations were quietly curtailed.
Even Dmitri{\u\i} Borisovich Zimin was forced to close his charitable foundation ``Dynasty'' in 2015.
Does anyone else continue to wonder about the ``brain drain'' from Russia?
Prying into the administration of science, without having an idea of how it functions, is -- to use Talleyrand's remark -- worse than a crime, this is a mistake.
Summing up the events of the 1990s in Russia, Zhores Alferov~\footnote{Zhores Ivanovich Alferov is a Soviet and Russian physicist, Winner of the Nobel Prize in Physics, academic of the USSR Academy of Sciences, Vice president of the Russian Academy of Sciences, Winner of the Lenin Prize, of the USSR State Prize and the Russian Federation State Prize, Full Cavalier of the Order of the
 Merit for the Fatherland.} stated: ``The preservation of scientific potential was partly successful due to participation in international scientific cooperation, international projects and grants''~\cite{Alferov}.
And I would add: we owe the survival of the Russian nuclear weapons centers at that time, to a large extent, to the 41st and 42nd US Presidents.

Of course, goodwill gestures and humanistic declarations in support of science in Russia from Washington do not gave rise to delusions: behind them cold interests were visible.
However, one should not go to the other extreme, transferring the hidden motives of the US administration to ordinary people.
Scientists, engineers, low-ranking officials involved in scientific cooperation were full of enthusiasm and idealistic mood.
I have managed three ISTC projects, contacted these people, and I know what I am talking about.
To help to {``turn the last page in the history of the Cold War and hostile relations between our countries''}~\footnote{I quote the wording, as I heard it more than once from my American colleagues, almost verbatim.} many of them considered it a matter of honor.
The page, unfortunately, remained unturned.

The reference to the presence of Russian nuclear centers is not capable of impressing the average American forever.
The lack of a national labs strategy also worries Washington.
Signs of anxiety can be seen even with the naked eye.
For example, over the past 30 years, the directors of LANL and LLNL have changed unusually often, the security services system has been reshaped, and the landscape of open research has changed dramatically.
For the first time in 60 years, LANL has changed its governing body.

There may have been a temptation to ``optimize costs'', that is, to reduce the scope of research and personnel not directly related to the problem of maintaining the combat capability of nuclear systems.
But in such a truncated world, isolated from big science, it is impossible to prevent a drop in the quality of research, a decrease in the qualifications of the employees of the National Laboratories to a level where they will no longer be able to critically evaluate their results.

Are the Russian nuclear centers and US National Laboratories, with their considerable budgets, really necessary?
If so, what is their purpose today and in the future?

I discussed these issues with Yuri{\u\i} Alekseevich Trutnev, Dmitri{\u\i} Vasil'evich Shirkov, Lev Vasil'evich Ovsyannikov, Nikola{\u\i} Aleksandrovich Dmitriev, authoritative veterans of the Atomic Project of the USSR; Isaac Markovich Khalatnikov, a major representative of theoretical physics in Russia at that time; Yuli{\u\i} Borisovich Khariton and Yevgeni{\u\i} Ivanovich Zababakhin, the scientific supervisors of, respectively, VNIIEF and VNIIP; Lev Dmitrievich Ryabev and Viktor Nikitovich Mikhailov, ex-ministers of the MSM~\footnote{The Ministry of Medium Machine Building  controlled the entire nuclear industry of the USSR since 1953.
After two renamings at the turn of the century, it became the State Atomic Energy Corporation Rosatom, or simply Rosatom, in 2007.}. 
Those were serious discussions.
Unfortunately, no single, strictly substantiated position has been reached.
Therefore, for all judgments, estimates and suggestions contained in this paper (as well as possible omissions and errors), I bear personal responsibility.

\subsection{Just like two little sisters from the years unlived, running onto the island, they will wave when I leave~\cite{Brodsky}}
\label
{systers}
Let us turn to some pages of the past.
By the beginning of the 1960s, most of the key figures of the Manhattan Project completed their activities in it \cite{Thorne}.
The same picture is in the Atomic Project of the USSR.
Mathematical leaders Nikola{\u\i} Nikolaevich Bogoliubov and Mikhail Alekseevich Lavrentiev are leaving VNIIEF with their followers.
The eminent physicists Yakov Borisovich Zel'dovich, Igor Evgenievich Tamm, David Albertovich Frank-Kamenetski{\u\i}, Isaak Yakovlevich Pomeranchuk, Lev Vladimirovich Altshuler are leaving.
Outside the ``object'', Israel Moiseevich Gelfand, Alexander Andreevich Samarski{\u\i}, Andre{\u\i} Nikolaevich Tikhonov, Vitali{\u\i} Lazarevich Ginzburg, as well as all project participants from 
the school of Lev Davidovich Landau, leave closed works.
The moment is approaching when Andre{\u\i} Dmitrievich Sakharov will announce: ``Farewell to arms!''.
And although the main principles of the operation of nuclear and thermonuclear weapons have been established, there will be a long period of its improvement, and systematic accumulation of knowledge about the processes of gas dynamics, the detonation of chemical explosives, nuclear chain reactions and thermonuclear fusion~\footnote{General impression of the results in fundamental physics obtained by theorists of VNIIEF, so to speak, in the intermissions between the quest for new nuclear devices, can be gained from the book \cite{Trutnev}.}.

Yuli{\u\i} Borisovich Khariton, Scientific Supervisor of VNIIEF~\footnote{Position similar to that held by Robert Oppenheimer at LANL.
But Oppenheimer was a theoretician, a disciple of the Niels Bohr school, while Khariton was an experimenter with qualifications acquired at the Cavendish Laboratory under the guidance of Ernest Rutherford.}, while developing a broad program for the investigations in nuclear physics, offers a brilliant young theorist from the Kurchatov Institute, Alfred Ivanovich Baz' \cite{BazUFN} to nurture for this purpose a theoretical group of $\sim 10$ best university graduates.
Remaining a member of the Kurchatov Institute, Baz' makes long visits to VNIIEF, where he does not touch on secret topics and has complete freedom of action.

For several years, Baz' founded a school of nuclear theory.
It seemed that the school was one step away from world recognition...
Alas, Alfred Ivanovich's life is cut short by an absurd accident.
The group of young theorists disintegrates shortly thereafter.
Some continue the activities begun under the leadership of Baz'; they reveal the structure of nuclei, and calculate their properties.
Others go to ``bomb-makers''.

The echo of those events reached me at the international conference {\it Mysteries, Puzzles and Paradoxes in Quantum Mechanics} held in Italy on Lake Garda in September 1999.
Many talks kept mentioning the Baz'--Rybachenko method.
I was familiar with this beautiful result~\footnote{The crux of the matter is this.
Let a particle with spin ${\bf s}$ and magnetic moment $2{{\mu}}{\bf s}$ tunnels through a potential barrier.
It is required to find the average time of its stay under the barrier ${\langle \tau\rangle}$ and other such quantities, for example, the variance ${({\Delta \tau})^2=\langle \tau^2\rangle-{ \langle \tau\rangle}^2}$.
In the region of interest to us, we turn on a weak magnetic field ${\bf B}$.
The spin will precess around ${\bf B}$ with the Larmor frequency $\omega=2{\mu}{\bf s}\cdot{\bf B}/\hbar$.
Therefore, ${\langle \tau\rangle}$ can be related to the angle of rotation of the spin in the region of action magnetic field.
Remarkably, the chronometric effects remain finite at ${\bf B}\to 0$.} obtained in 1966--1967, \cite{Baz}, \cite{Rybachenko}, but did not think that it would be so popular in the West after more than 33 years.
On the {couloirs}, I learned that the military wants to have quantum computers and quantum cryptography at their disposal.
Obviously, such problems require a deep understanding of the conceptual foundations of quantum mechanics.
Therefore, for research in this area, including the most delicate and expensive experiments~\footnote{While these lines were being written, the news came that the Nobel Prize in Physics in 2022 was awarded to Alain Aspect, John Clauser and Anton Zeilinger for experiments with entangled photons, establishing the principle of violation of Bell's inequalities, and for discoveries in the field of quantum information science, {\it i. e.}, experiments aimed at clarifying the foundations of quantum mechanics.}, they spare no expense.
As for the Baz'--Rybachenko method, the situation here is slightly different.
It is directly relevant to modern technologies in electronics, where the size of microcircuits is already approaching the atomic scale, so that the life of a particle under barriers becomes the concern of other wealthy patrons -- from the civil sector.

A large delegation of young, full of enthusiasm, researchers of the microworld arrived in Italy from LANL.
Once, in a conversation with them, I dropped that I work at the same institute with Vladimir Fyodorovich Rybachenko.

``Is he alive?'' they asked me.

``Alive, healthy,'' I replied,  ``and well fed.'' 

``But we know that Zel'dovich and Baz' are dead. In the USA people think that Rybachenko also dead,'' they continued, ``Maybe he was in prison, like Landau?''

``No. Why? He did not call for the overthrow of power in the country.''

``Then why did he stop publishing his papers so suddenly?''

``I don't know. When I get home, I'll ask him.''

Returning to Sarov, I asked Rybachenko to tell what happened to him then.
That is what I heard.
Volodya was 27 years old.
He succeed in publishing about a dozen good papers in good journals.
But what he wanted most of all, rushing along in the stream of research organized by Baz', was to define the time operator ${\hat t}$ in quantum mechanics.
Zel'dovich was intrigued by this idea (after all, this challenge was beyond the power of the founding fathers of quantum mechanics) pushing Baz' and Rybachenko to sum up their results in a review article, for which Zel'dovich had already reserved a place in the portfolio of forthcoming publications of the Soviet Physics--Uspekhi, the central physical journal of the USSR.
The draft of the article was ready.
But something went wrong in Baz's life; he lost interest in the idea of the ${\hat t}$ operator.
Volodya did not dare to complete the project on his own.
At that time, Zel'dovich was completely absorbed in the affairs of the school of astrophysics and cosmology, which he had recently founded.
It was the world's strongest gravitational team \cite{Thorne}, and leadership demanded extreme effort from the leader \cite{Zeldovich}.

Volodya did not find anything better than to turn to Sakharov for moral support.
Sakharov was leaving VNIIEF, and this was his last visit to the ``object''.
An official car was sent to the station for Sakharov.
It was a sunny May morning.
Sakharov invited Volodya to take a ride in the blue ``Volga'' to the Institute, and on the way state your question.
Everything worked out very well.
Volodya spoke with inspiration.
Sakharov listened in silence, then closed his eyes and apparently fell asleep.
Volodya was embarrassedly silent.
Then Sakharov opened his eyes, and in a soft tone, briefly but decisively, advised Volodya give up scholasticism and do something more useful.
Sakharov, no doubt, realized that his personal example would be incomparably stronger than any instructions, and that he himself, moving away from practical work at the ``object'' in favor of studying academic sciences, deprives his advice of persuasiveness.
On the other hand, it was clear to him that before him was a man who was hardly ready to stand alone before the blows with which big science would not be slow to fall upon him.
And, therefore, my friend, do not ruin your life -- do not retreat far from the interests of your team.
Sakharov was right.
Volodya carefully collected his publications, calling this anthology his Ph.D. 
thesis, successfully defended it, said goodbye to nuclear theory and other more or less abstract research, and devoted his talent entirely to ``bomb-making''.

\section{Nuclei are emeralds pure, but can I of this be sure?~\cite{Pushkin}}
\label
{Kernels}
Nuclear physics represented the frontier of physical science for about half a century.
In 1911, Rutherford discovered atomic nuclei and proposed the very term ``nucleus'', but already in 1954, the world's first nuclear power plant was put into operation in Obninsk, and in 1958, Aage Bohr, Ben Mottelson, and David Pines invented the superfluid model of nuclei.
Then attention to the nucleus quickly faded.
At present, nuclear physics has been relegated to the periphery of physical research.
At least in the Nobel Committee, interest in its achievements seems to be completely dried up.

The application of nuclear physics resulted in unprecedented in scale accomplishments: the atomic fleet,  nuclear power plants, and nuclear weapons.
And yet, does this mean that we understand the physics of nuclei and subnuclear realm to such an extent that we do not feel any concern for its applications and the fate of its theoretical foundations?
I will try to answer this question, but doing this I will be forced to cut off many corners and omit important technical details.

\subsection{Look at the root~\cite{Prutkov}}
\label
{inside}
Since the 1950s, the frontier research has been dominated by quantum field theory and particle physics~\footnote{In modern parlance, these are, respectively, the theoretical and phenomenological branches of High Energy Physics (HEP).}.
The work of many talented physicists contributed to these activities.
However, the history of subnuclear physics contains not only victorious pages.
The first decade-long crisis occurred in the 1960s.

The success of using perturbation theory in quantum electrodynamics \cite{Schweber} stimulated attempts to apply it to the Yukawa model of strong interactions and the four-fermion model of weak interactions.
The difficulty was that the coupling constants of the strong interaction of hadrons exceed 1 in absolute value, and therefore, in contrast to the fine structure constant $\alpha\approx 1/137$, they are unsuitable for the role of small parameters in perturbation theory.
As to four-fermion interactions, the renormalization method proved to be in principle unfit for this model.

To top it off, Landau and his colleagues got a stunning result, called ``zero-charge'' in the USSR , and ``Moscow zero'' in the West.
The essence of it is as follows.
Around any charged point particle, quantum fluctuations give rise to pairs of virtual particles and antiparticles.
Lined up like dipoles, the pairs screen the initial charge.
This phenomenon, characteristic of relativistic quantum theory, is called {\it vacuum polarization}.
The authors of Refs. \cite{Landau-} and \cite{Fradkin} concluded that the vacuum polarization in renormalizable theories is so great that the ``bare'' charge is completely screened, in other words, the interaction between particles with renormalized coupling constants is switched off.
Another difficulty related to the zero-charge problem is the ``ghost'' state of the photon \cite{Abrikosov}.

Landau believed that the nullification of the charge means the logical inconsistency of the renormalizable theories, implying the collapse of the concept of local interactions \cite{Landau}.
In the West, this verdict was taken quite seriously: ``Under the influence of Landau and Pomeranchuk,  a  generation of physicists was forbidden to work on field theory'' \cite{Gross}.
In the United States, theorists began to study the interactions of particles using the analytical theory of  $S$ matrix, unitary symmetries, and current algebra.
And what were the further steps of theorists in the USSR?

Pomeranchuk closed the weekly seminar on field theory at ITEP, but two months later resumed its work, switching the attention of its participants to phenomenological descriptions of scattering amplitudes that are not related to the Feynman technique in perturbation theory.
Research in this direction has yielded many interesting results, in particular Pomeranchuk's theorem on the equality of the total cross sections for the interaction of particles $P$ and antiparticles $\bar P$ with a target $T$ in the limit that the energies of $P$ and $ \bar P$ tend to $\infty$ \cite{Pomeranchuk}.
The vocabulary of particles has been enriched by the term ``Pomeron'', denoting a micro-object with vacuum quantum numbers.
 
On the other hand, Bogoliubov found the mourning for quantum fields  inappropriate.
Indeed, the zero-charge problem arose as a result of summing some set of terms of the perturbation theory series in powers of $\alpha$.
Freeman Dyson established \cite{Dyson} that such a series is divergent, so it is to be only taken as an asymptotic series.
There is well-known techniques for summing divergent series.
However, the results of such operations give not a single-valued analytical expression but rather a set of functions that differ from one another by an entire function ({\it i. e.}, a function that is holomorphic everywhere in any finite part of the complex plane of the parameter $\alpha$).

Getting into the spotlight of theorists of scattering amplitudes that behave as entire functions gives a signal to storm nonlocal, essentially nonlinear, and nonrenormalizable field theories.
In the 60s and 70s, the attack was a success.
A strict line of demarcation between localizable and nonlocal field theories \cite{Meiman}--\cite{Efimov68} has been defined.
It was found out how to construct the $S$ matrix for interacting fields with nonlocal form factors which is free from ultraviolet divergences and satisfies all the general conditions of quantum field theory: Lorentz covariance, unitarity, and causality \cite{Efimov}--\cite{AlebastrovC}.
A generalization of the most important results of axiomatic quantum field theory (spin-statistics theorem, $PCT$ invariance) to the case that the vacuum expectation values of quantum fields grow exponentially with energy, which is typical for nonlocal interactions of quantized fields, has been gained \cite{IofaFainberg}.
Techniques for dealing with nonpolynomial Lagrangians \cite{Fradkin63}--\cite{Volkov} were invented.

Nevertheless, the main direction in understanding the three fundamental interactions: weak, electromagnetic, and strong,  as the natural course of events showed, was related to local renormalizable quantum field theory.
Its scope is noticeably narrowed.
Acceptable forms of interaction include only those which owe their origin to the {\it Yang--Mills field} theory \cite{YM, Utiyama}.
What did there serve as a guiding star?
First of all, the inspiring case of quantum electrodynamics~\footnote{This case was both inspiring and discouraging.
Indeed, the null-charge problem concerns a conceivable exact solution of local quantum field theory.
Strange though it may seem, if we are interested in an approximate solution, then quantum electrodynamics provides us with both well-developed scheme for finding it and excellent agreement between theoretically calculated and experimentally measured effects.
The paradox, then, is that the theory as a whole is regarded as logically inconsistent, while its approximate solutions are physically perfect.}, where the interaction carries a gauge vector field.
Chiral symmetry of the strong interactions and the vector-pseudovector (more precisely, $V-A$) nature of the weak interactions suggested that particles with spin 1 could be the carriers of these interactions.
Hence, the close attention to the vector fields of gauge theories as agents of the strong and weak interactions.

The unification of the electromagnetic and weak interactions occurs on the basis of the gauge group ${\rm SU(2)}\times{\rm U(1)}$ \cite{Glashow, SalamWard}, which implies the presence of vector bosons $A_\mu$, $W ^{\pm}_\mu$, and $Z_\mu$.
As far as the strong interactions are concerned, here {\it quarks} came to the fore as elementary constituents of hadrons \cite{Gell-Mann, Zweig}.

However, these findings were not taken seriously.
First, in the early 1960s, we still did not know how to quantize gauge theories.
Secondly, although massless Yang--Mills fields are suitable as carriers of long-range forces like photons responsible for the electromagnetic interaction, to carry the weak interactions limited by distances in the range of $\sim 10^{-16}$ cm, one needs massive vector bosons.
Thirdly, if we try to modify the Yang--Mills theory by simply supplying the gauge fields $W^{\pm}_\mu$ and $Z_\mu$ with masses, then the resulting theory turns out to be nonrenormalizable.
Fourthly, there was no way to extract quarks from hadrons and experimentally study their properties in an isolated state \cite{Lyons}.
Therefore, Murray Gell-Mann himself, who proposed the idea of quarks, treated them as mathematically convenient objects, which, however, do not claim the status of real particles.
And, finally, gauge theories suffered from quantum anomalies \cite{Adler, BellJackiw}, the presence of which threatened the $R$-operation (the only tool of renormalizable theories \cite{Bogoliubov}--\cite{Zimmerman}) with deprivation of its working quality.
 
Quantization of Yang--Mills fields that does not violate gauge invariance and unitarity \cite{FaddeevPopov} led to the birth of bizarre entities, the {\it Faddeev--Popov ghosts}.
They are fields that do not obey the spin-statistics rule.
The ghosts live only in virtual states, namely in loops of Feynman diagrams.
There are no real particles corresponding to quanta of these fields.

The possibility of endowing gauge bosons with masses without affecting the properties of renormalizability has been made possible to find once a more penetrating insight into the concept of {\it spontaneous symmetry breaking} \cite{Goldstone} and the {\it Higgs mechanism} \cite{Englert}--\cite{Higgs}, which allows to preserve gauge invariance even if some interaction-carrying 
Yang--Mills fields acquire mass, has been gained.

With these discoveries, a realistic model of the electroweak interaction was proposed, in which the ${\rm SU(2)}_L\times{\rm U(1)}_Y$ symmetry group was spontaneously broken down to the electrodynamics gauge group ${\rm U(1 )}_{\rm em}$ \cite{Weinberg67}, \cite{Salam}.
The renormalizability of this model was rigorously established 5 years later \cite{HooftVeltman}.
In parallel, the idea of {\it color degrees of freedom} \cite{Greenberg}--\cite{BST} was developed, which formed the basis of gauge theory for the strong interactions with the unbroken color gauge symmetry group, ${\rm SU(3)}_c$ \cite{Weinberg73}, \cite{Fritzsch}.
This theory, formulated in terms of quarks and gluons, is now called {\it Quantum Chromodynamics}~\footnote{$X\!\varrho {\acute \omega} \mu \alpha$ is Greek  for color.} (QCD).

The most prominent feature of QCD is the {\it asymptotic freedom} phenomenon.
It was found in Refs. \cite{GrossWilczek, Politzer}~\footnote{In fact, the same result was obtained in the USSR a little earlier in Refs. \cite{VanyashinTerent'ev, Khriplovich}, but those findings did not properly assessed by their authors.} that the renormalized coupling constant weakens logarithmically with increasing the energy renormalization scale, $\alpha_s(E)\propto\left(\ln E\right)^{-1},\, E\to\infty$, in other words, if the distance between quarks tends to zero, the color coupling between them turns off.
This explained the mysterious angular behavior features of the cross sections for deep inelastic lepton-hadron collisions \cite{Feynman}.

Technically, a favorable consequence of asymptotic freedom is that the QCD coupling constant, $\alpha_s(E)$, becomes small for sufficiently large values of energy $E$~\footnote{It is more properly to refer to the four-momentum transfer squared rather than energy.}.
Therefore, it can be used as a small parameter in the perturbation theory to calculate higher corrections in amplitudes that describe quark-gluon and gluon-gluon interactions.

From the standpoint of the foundations of subnuclear physics, QCD might appear to be a counterexample to the zero-charge doctrine.
In contrast to quantum electrodynamics, where the vacuum polarization screens the bare charge, in QCD there is a competition between screening and anti-screening effects due to the fact that gluons have color, and can interact with each other.
If the number of quark species $n_f$ is not too many ($n_f\le 16$), then antiscreening wins.
However, the ``ghost'' pole did not disappear from the gluon propagator, but simply moved from the ultraviolet to the infrared.
The question of logical consistency remains in QCD as well.

A remarkable fact is that if $n_f$ is equal to the number of lepton species, then the quantum axial anomalies cancel \cite{GrossJackiw, Bouchiat}.
Presently, 6 species of quarks, taken two at a time: $(u, d)$, $(c, s)$, $(t, b)$, and 6 species of leptons: $(e, \nu_e)$, $(\mu, \nu_\mu)$, $(\tau, \nu_\tau)$ have been found, and there is no reason to expect this situation to change in the future.
The anomaly cancellation condition requiring the ``quark-lepton symmetry'', played a key role in determining the composition of matter with the {\it three generations} of quarks and leptons.

This is how the {\it Standard Model} (SM) of elementary particles was made step by step \cite{Goillard}, \cite{Okun}.
Then the triumphal process of its experimental confirmations starts, the most important stages of which are the discoveries of $c$, $b$, and $t$ quarks, $\tau$-lepton, vector $W^\pm$ and $Z$-bosons, and the Higgs scalar boson.
But the experimental picture left a strange impression, because its content continued to be filled with direct measurements in the dynamics of hadrons, not of quarks and gluons.
To explain the absence of free quarks and gluons, the {\it confinement} hypothesis of color objects in hadrons was put forward.

All observed states of systems of color objects must form singlets of the ${\rm SU(3)}_c$ group, in other words, be colorless \cite{KugoOjima}--\cite{Bander}.
The fate of being locked up forever did not seem quite unthinkable for color objects after the Faddeev--Popov ghosts received citizenship in the subnuclear worlds.
The confinement hypothesis is supported by the increase of the QCD coupling constant $\alpha_s(E)$ in the infrared limit, {\it i. e.}, at the range of energies $E$ that are noticeably lower than the characteristic QCD energy, $\Lambda_s\sim 100$ MeV, or, which is the same, at distances much larger than $r\sim 10^{-13}$ cm.
Admittedly, for $\alpha_s\ge 1$ calculations using the perturbation theory are impossible, so the above argument may be only a qualitative indication of the implementation of confinement. 
Another argument is provided by the results of calculations on lattices \cite{Wilson74}, \cite{Creutz}.
It is far from easy to prove the confinement hypothesis rigorously.
This is part of the solution to the four-dimensional quantum Yang--Mills problem, one of the seven {\it Millennium Problems} announced by the Clay Mathematical Institute \cite{JaffeWitten}.

Let us return for a while to the 60s--70s, to the attack on nonlocal, nonrenormalizable, and essentially nonlinear field theories.
Where did that fever pitch developed into?
During that period,  a clear strategy for finding exact solutions to quantum field theories was outlined.
Exact solutions of four-dimensional quantum models~\footnote{There are many exact solutions to {\it classical} Yang--Mills theory \cite{Actor}.
However, their discovery was not followed by experimental discoveries.
The 't Hooft--Polyakov monopole \cite{Hooft}, \cite{Polyakov} has existed for half a century only on paper.
The instanton \cite{Belavin} did not give a clue to the confinement mechanism.
It is interesting that some inhabitants of the microworld do not have a pictorial rendition associated with a solution of classical field equations.
Therefore, nothing can be said about their characteristic sizes and configurations.
Sydney Coleman established \cite{Coleman} that such an object is glueball, a colorless bound state of two or three gluons.}, alas, are still missing.

The Monte Carlo method in the Euclidean lattice formulation of gauge theories has gained great popularity in the search for exact QCD solutions \cite{Briceno}.
However, this approach is applicable only to elastic processes, {\it i. e.}, it is limited to energies below the threshold for multiple particle production.
In addition, it is not clear whether the result of calculations on Euclidean lattices can be analytically extended to the four-dimensional Minkowski continuum.
There are no sensational breakthroughs here yet.
Note that even attempts to reformulate the SM in lattice terms fared poorly.

Are exact solutions really necessary?
Why not limit ourselves to approximate solutions if they are in good agreement with experimental measurements?
I could here simply refer to the lesson taught by the zero-charge problem.
However, it is more important to note that the answers to deep questions give exact solutions.

Consider, for example, the stability problem for a system of two particles with charges of opposite signs.
If we restrict ourselves to a static classical picture, then the system is found to be unstable.
In the framework of Newtonian dynamics, the centrifugal force is able to give the system stability.
However, if the effect of relativistic retardation in the propagation of electromagnetic interaction is taken into account, the system again turn out to be unstable \cite{Synge}.
And considering the energy loss due to the radiation of accelerated charges, we all the more come to the conclusion that the fall of both particles on the center of mass is inevitable.
At the beginning of the last century, this conclusion led to a crisis in the classical description of stable matter, and this was one of the main reasons for the advent of quantum theory.
In Bohr's and Schr\"odinger's versions of quantum mechanics, the system regains stability \cite{DysonLenardI}--\cite{DysonLenardII}.
What will exact solutions to the relativistic quantum field equations say about stability?
The experiment does not give clues.
Indeed, every hydrogen atom is stable on Earth, but once it hits a neutron star, it collapses and becomes a neutron with the emission of an antineutrino.

Let us continue the particle saga.
In the early 70s, when it was still very far from the proper order in affairs of the SM, theorists burned with impatience to rush along the energy scale up to the Planck limit $E_{\rm Pl}=\sqrt{\hbar c^5/G_N}\approx 1.2\cdot 10^{19}$ GeV, only stopping at the point of {\it Grand Unification} (GU) of the weak, electromagnetic, and strong interactions, $E_{\rm GU}\sim 10^{16}$ GeV.
Above this energy, leptons and quarks turn into components of a single fermion field; and intermediate bosons $W$ and $Z$ become massless, like gluons and photons, forming together with vector fields $X$ and $Y$ components of a single gauge field (everything is almost according to Isaiah: ``And the lion shall eat straw like the ox'').

The quest for GU models \cite{PatiSalam}--\cite{Langacker} did without revolutionary shocks.
The symmetry group ${\rm SU}(2)\times{\rm U(1)}\times{\rm SU}_c(3)$ should be embedded in a semisimple group, say SU(5) \cite{GeorgyGlashow}.
The dynamics of GU models relates to the SM dynamics through the renormalization group.
At large $E$, the reciprocals of the running couplings of the weak, electromagnetic, and strong interactions, $\alpha_w^{-1}(E)$, $\alpha^{-1}(E)$, and $\alpha_s^{-1}(E)$, are proportional to $\ln E$, although the coefficients of proportionality are different.
If all three lines converge to one point, then it is reasonable to refer to the coordinate of their meeting as the Great Unification energy $E_{\rm GU}$.
 
Proton decay is the most striking phenomenological prediction of GU theories.
Indeed, since quarks and leptons are interconverted, the baryon and lepton numbers are not conserved quantities.
The simplest proton decay scheme $p\to e^++\pi^0$, admissible in the minimal GU model with the SU(5) group, is that left and right $u$ quarks entering into the composition of the proton, annihilate into $X^+$ boson that decays into $e^+$ and ${\bar d}$ quark, and the latter together with the ${d}$ quark of the proton form $\pi^0$.
According to this mechanism, proton decay occurs after $\sim 10^{31}$ years.
Proton decay, with converting a quark into a muon, was discussed by Sakharov \cite{Sakharov67} long before the emergence of the GU models, when neither energy $E\ge M_{X^+}\approx 2.5\cdot 10^{14}$ GeV nor coupling constant $\alpha_{\rm GU}(M_{X^+}^2)\approx 0.024$ were yet known.
Reasoning from several arbitrary assumptions about such quantities, Sakharov estimated the proton lifetime at $\sim 10^{50}$ years.

This prediction could unlikely kindle enthusiasm among experimenters.
Another thing is checking the instability of a proton with a lifetime of $\sim 10^{31}$ years.
A ton of water contains $\approx 3\cdot 10^{28}$ protons.
Therefore, in a ball of water with radius of $\ge 10$ m, one can expect several decays of protons per year.
The ball should be wrapped around with several thousand photomultipliers and lowered as deep as possible underground to cut off the cosmic ray background.
A dozen of such units were ready for operation in the mid-80s.
But not a single case of decay has been recorded on them.
The experiment continues only on the Super-Kamiokande set up with the mass of water of 50,000 tons, compatible with the limiting level of sensitivity of the measuring equipment, and, as before, does not give encouraging results.
According to current data, the lower limit of the proton lifetime in this channel is $1.67\cdot 10^{34}$ years \cite{Zyla}.

The search for proton decay is an ambitious experimental project.
Great hopes were pinned on him.
By the early 1990s, it became clear that the project had failed.
There was a breath of cold from the impending decline of high energy physics.

In 1971, Soviet and Western physicists invented {\it supersymmetry}, namely symmetry between fermionic and bosonic degrees of freedom~\cite{BerezinKats}--\cite{VolkovAkulov}.
In supersymmetrized Yang--Mills theories ultraviolet divergences were found to cancel in the one-loop approximation.
Some of these theories have proved to be finite in all orders of perturbation theory.
The hope arose for the construction of a finite theory of all four fundamental forces of nature.
The initial version of this theory, called {\it supergravity}, was a supersymmetric combination of Grand Unification models and General Relativity~\cite{VanNieuwenhuizen}.
However, neither the titanic efforts nor the miracles of the theoreticians' inventiveness were crowned with success.
The divergences in supergravity are actually irremovable.

In 1968, Gabriele Veneziano adapted the Euler $B$-function to describe the scattering amplitude of two particles in the dual resonance model~\cite{Veneziano}.
The idea was immediately picked up by researchers involved in the phenomenology of strong interactions.
It soon became clear that this amplitude encoded the dynamics of a relativistic {\it string} with an action proportional to the area of world surface swept by this string in the course of its evolution~\cite{Nambu}, \cite{Goto}.
Extended objects, strings, settled in the picture of the microworld.

However,  consistent quantizations of bosonic strings was found to be possible only in 26-dimensional spacetimes; and only 10-dimensional manifolds are suited for quantizations of superstrings.
In addition, the spectrum of closed strings contains excitations with spin $2$, and no particles with such a spin have been observed among hadrons.
All this cooled interest in the string description of strong interactions.

After a lapse of 5 years, strong interactions acquire a quantum field basis, QCD.
The strings are given a new role.
They become the objects responsible for all the fundamental forces of nature~\cite{ScherkSchwarz}.
To relate the superstring realm to the observed world, the six extra dimensions are curled up into a tiny tube following the Kaluza--Klein {\it compactification} prescription.
The spin-2 string excitations are treated as gravitons.
The behavior of strings is governed by a two-dimensional conformal quantum field theory.
Joinings and splittings are the only form of interaction of strings in a perturbative framework.
It is assumed that the characteristic size of strings is comparable to Planck's length ${\ell}_{\rm Pl}= \sqrt{\hbar G_N/c^3}\approx 1.6\cdot 10^{-33}$ cm, and the compactification scale is also comparable to ${\ell}_{\rm Pl}$.
 
Three {superstring revolutions} then befell High Energy Physics.
About two of them the reader may consult the textbooks~\cite{GreenSchwarzWitten} and~\cite{Polchinski}.
The first revolution was able to cope with  quantum anomalies to result that only 5 types of string theories were taken to be acceptable: type ${\rm I}$, type ${\rm IIA}$, type ${\rm IIB}$, and two heterotic theories with SO(32) and ${\rm E_8}\times {\rm E_8}$ symmetry groups.
The type-${\rm II}$ theories have two supersymmetries in the  10-dimensional sense, while the other three have just one.
The type-${\rm I}$ theory describes open and closed strings; an open string can become closed or split into two strings, and these processes are reversible.
In other theories, the strings are always closed.

Strings are immune from the ultraviolet disease.
The supersymmetric cancellation of divergences bears no relation to it because the theory of bosonic strings is also finite~\footnote{Note, however, that there is a tachyon in the spectrum of bosonic strings, which can be removed with the help of supersymmetry.}.
It is still common the assertion that these theories are finite due to the extent of the strings.
This statement is wrong, as exemplified by another extended object, a two-dimensional membrane with an action proportional to its world volume ${\cal V}_3$.
In an effort to build a local quantum field theory on ${\cal V}_3$, you find yourself in a gloomy forest of ultraviolet divergences.

The compactification of ${\rm E_8}\times{\rm E_8}$ heterotic strings on the Calabi--Yau manifolds and orbifolds yields a consistent theory called the {\it Theory of Everything}.
In the low energy limit ($E\ll 10^{19}$ GeV) it contains classical supergravity and Grand Unification models with chiral representations for quarks and leptons.

We thus have at our disposal 5 versions of a consistent theory.
Which one should be preferred?
The compactification only exacerbates the difficulty of choice.
There are a huge number of Calabi--Yau configurations describing vacuum states with the same energy, and it is not clear how to narrow the set of acceptable theories.

The second string revolution showed that they all are in fact manifestations of the same physics in different contexts, for example, in regimes of strong and weak couplings.
Any two of these theories are related by a duality transformation \cite{Polchinski96}.
Superstrings and soliton-like objects of supergravity are tightly entangled in a web of dualities to give a single nonperturbative scheme called {\it $M$-theory} \cite{Witten95}.

A rigorous proof of the dualities will be possible after finding nonperturbative solutions of the linked theories.
However, the decoding of dynamical systems in the strong-coupling regime is still beyond our capabilities.
Therefore, many dualities remain mere plausible conjectures.

One of the features of $M$-theory is the existence of extended entities with different number of dimensions, and these entities are readily interconverted.
However, it is not clear which degrees of freedom are fundamental in $M$-theory. 
In any case, those are not particles, not strings, and not branes (objects of two spatial dimensions and more).

Due to its speculative nature, complexity, and lack of predictions, interest in $M$-theory quickly cooled off.
We cannot use the finiteness of string theories to calculate processes with quarks and leptons at available energies, because we do not know the veritable mechanism of descent from 10 to 4 dimensions.
For comparison, recall that classical relativistic mechanics allows us to calculate the motion of particles with both velocities close to the speed of light and low velocities, and often such relativistic calculations are simpler than those in Newtonian mechanics.

The third string revolution declared the presence of a holographic mapping of string physics onto physics of phenomena that occur at the energy range provided by present-day accelerators \cite{Maldacena}--\cite{Gubser}.
This approach is called {\it gauge/gravity duality} and {\it correspondence between gravitation in anti-de Sitter space and conformal field theory}, AdS/CFT.
Ideas and techniques of this approach are presented systematically in~\cite{Ammon} and \cite{Nastase}.

The holographic principle is the assumption that part of the subnuclear physics in our 4D world is patterned after the physics of black holes and similar objects (black branes, black rings, etc,) in a 5D anti-de Sitter space, ${\rm AdS}_5$, of which this 4D world is the boundary~\footnote{The origin of AdS/CFT holography is explained as follows. 
The manifold ${\rm AdS}_5$ can be thought of as a five-dimensional hyperboloid in a six-dimensional pseudo-Euclidean space ${\mathbb R}_{2,4}$. 
This hyperboloid is invariant under the orthogonal group ${\rm SO}(2,4)$.  
But the group ${\rm SO}(2,4)$ is isomorphic to the conformal transformation group C(1,3) acting on the four-dimensional Minkowski space ${\mathbb R}_{1,3}$. 
Thus, the geometry of the ${\rm AdS}_5$ manifold itself induces a conformal symmetry of its holographic image.}.
Knowing the solutions of classical gravity in ${\rm AdS}_5$~\footnote{Classical gravity is a legacy of superstrings on manifolds like ${\rm S}^5\times{\rm AdS}_5$ in the low-energy limit.}, we are trying to understand what happens in QCD in the strong-coupling regime.
For example, a Schwarzschild black hole in ${\rm AdS}_5$ is holographically mapped into a lump of quark-gluon plasma on a 4D screen~\cite{Herzog}.
For a quarter of a century, a large body of research on the gauge/gravity duality has been published with a pletora of explanations of microworld phenomena.
But not a single experimentally confirmed prediction has been proposed.
The predictive power of holography is still not known.

And here is the denouement of the tragedy.
The notion of  particles become unnecessary in quantum gravity~\cite{Witten22}.
This is no great surprise.
In quantum field theory, a particle is understood as an excitation over the lowest energy state in the Hilbert space of states.
But the very concept of energy as a conserved quantity in an arbitrarily curved spacetime cannot be defined.
Therefore, the concept of a particle is also meaningless here.

By the end of the 20th century, the theoretical part of HEP had become so divorced from its experimental part that the problems of the sub-Planck region (and even the GU region) are often taken to be unrelated to theoretical physics, and regarded as the subject of speculative constructions within the framework of abstract mathematics \cite{Woit}.

\subsection{Ne sutor supra crepidam judicet~\cite{Plinius}}
\label
{ne sutor supra crepidam judicet}
The heroic breakthrough in High Energy Physics was crowned with establishing the SM, GU models, supergravity, superstrings, $M$-theory,  loop gravity \cite{Rovelli}, and gauge/gravity duality.
However, beyond the limits of applicability of the SM, we are treading on shaky ground.
The search for proton decay did not give a positive result.
At the beginning of the 21st century, the victorious march into the subnuclear realm came to the end.

This, of course, is not a reason for despondency and agnostic lamentations ``ignoramus et ignorabimus''~\footnote{From Latin: We don't know and we won't know.}.
Indeed, in no foreseeable future will we be able to travel outside the solar system, but it does not follow from this that we will not be able to obtain reliable knowledge about distant worlds.
And then, psychologically, we are reassured by the fact that great discoveries await us inside the solar system.

Equally, it is worth recognizing that there are still many blank spots in the SM, where not only ``why'', but also ``how'' is still inappropriate.
For example, $u$ and $d$ quarks come in two forms: ``current'' and ``constituent''.
For the remaining species of quarks, the demarcation line between current and constituent mysteriously disappears.
Physicists rarely intrude into the field of epistemology, asking: ``Why is this so?''
But even a modest technical question: ``How to calculate the reaction $d\to u+ e +{\bar \nu}_e$?''
does not have a satisfactory answer today.
The calculation of measurable quantities for weak decays is possible only in simple bound systems of quarks, as exemplified by $\beta$-decay of a {free} neutron.
We learn the probability of $\beta$-decay of nuclei from the experiment.

The explanation of the quark coupling mechanism in QCD does not extend beyond systems with two and three quarks.
It is based on the dual Meissner effect, namely, the QCD vacuum is likened to a superconducting medium where the weak color field is completely displaced, while the strong one continues to exist in vortices that penetrate the bulk of the superconductor.
The flow of chromoelectric lines between a quark and an antiquark forming a meson is squeezed into a string.
This ensures the keeping of quarks together because the force of mutual attraction of quarks does not depend on the distance between them.
The configuration as a whole resembles a dumbbell~\footnote{This rendition of a meson can be verified experimentally if we have at our disposal a sufficiently strong magnetic field or strong plane wave electromagnetic field.
The quark and antiquark that form the $\pi^0$ meson have opposite electric charges.
In a magnetic field, they move in circles in different directions.
Therefore, the dumbbell can stretch, the wave functions of the quarks canl begin to overlap progressively less, the probability of their annihilation, $\pi^0\to\gamma\gamma$, can decrease, the lifetime of the $\pi^0$ meson can increase.
Further stretching of the dumbbell will lead to a break of the color string, {\it i. e.} the number of $\pi^0$ mesons will increase.
Note that $\pi^0$ is a truly neutral particle, so its cloning is in complete agreement with the principles of quantum field theory.
To carry out an experiment of this kind, fields with a strength close to the Schwinger limit will be needed.
The possibility of creating laser fields of such intensity is discussed in Ref. \cite{Bulanov}.}.
For a three-quark system connected by color strings to result in the formation of a baryon, two types of configurations are possible: a triangle in the corners of which quarks are located, and a star with three rays at the ends of which quarks are suspended.

Nucleons are colorless three-quark objects.
It is customary to think that nucleons are bound in nuclei by residual color forces like the van der Waals forces between neutral molecules.
However, this is a mere qualitative consideration.
I have not come across specific analytical calculations of the residual color multipole interaction resembling the Yukawa attraction $-g^2e^{-m_\pi r}/r$.
So, is the idea of residual color forces true at all?
An alternative point of view is to abandon the model of the nucleus as a system of ${\cal A}$ nucleons bound together by the exchange of mesons.
One can imagine the nucleus as a collection of $3{\cal A}$ quarks not forced to live in triple rooms.
Indeed, a free neutron, when combined with a proton to form a deuterium nucleus, loses its individuality.
This can be seen at least from the fact that the lifetime relative to $\beta$-decay increases from $T\approx 15$ minutes to $T=\infty$, {that is}, the neutron ceases to be responsible for the fate of its quarks.
The structure of the nucleus as a ``kibbutz'' of $3{\cal A}$ quarks will require rethinking the problem of confinement in a broader context that includes nuclear physics.

Rapidly moving into the depths of the microcosm, the theory left the affair of  nuclear physics in the form even more mysterious than that in the 30s.
The gluon field, like the electromagnetic field, is massless, so the color forces are long-range.
But electromagnetic forces allow the formation of stable molecules with any number of atoms, while the mass number of stable nuclei is limited by the value ${\cal A}=208$, corresponding to lead.
What is so special about the number 208?
There are also giant systems of quarks, neutron stars, but stability in them is provided not by the laws of QCD, but rather by the balance between the gravitational compression of a large mass of a star and the resistance to this compression, due to the pressure of quantum degeneracy.

We are far from understanding the structure of the Mendeleev periodic table.
Light stable nuclei up to ${\cal A}=40$ consist of equal parts of protons and neutrons, $Z=N$.
More precisely, among stable isotopes of a given element, there will always be one for which this equality holds, with the exception of beryllium, which is represented by the only stable isotope with $N/Z=1.25$.
For ${\cal A}>40$, stable nuclei become neutron-rich, and the excess of neutrons grows linearly with $Z$ up to $Z=82$.
Then nuclei cease to be stable at all.
Something prevents the formation of stable nuclei from neutrons alone~\footnote{In QCD terms, stable systems with an equal number of $u$ and $d$ quarks are forbidden.}.
We need to find an explanation for these facts, if not from the first principles of QCD, but at least within the framework of an effective theory to low-energy QCD.

When on the subject of theoretical essentials of high energy physics, we can confidently rely only on the axiomatic quantum field theory (summarized in the books \cite{StreaterWightman}--\cite{Haag}).
It is interesting to understand why the rigorous results of quantum field theory remain valid if the system of axioms expands its framework to cover nonlocal interaction of fields, but this does not happen when trying to extend the arena of events by replacing Minkowski spacetime with curved manifolds.
Does this mean, for example, that taking into account gravity, the spin-statistics theorem ceases to hold in some sense?

The theory of gravity itself will probably also be revised.
The idea of gravity as a curvature of space, dating back to Carl Friedrich Gauss, Bernhard Riemann, William Clifford, and Henri Poincar\'e, was embodied in the general theory of relativity (GR) by Albert Einstein and David Hilbert.
An important point is that Poincar\'e believed that the choice of geometry is a matter of agreement.
The true geometry of the physical world does not exist.
The geometry is chosen for reasons of convenience.
Poincar\'e himself considered Euclidean geometry to be the most convenient.
One can take a different description of the properties of spacetime if the physical laws are properly changed; as a result, the set of physical events and relations between them will remain the same \cite{Poincare}.
This tenet is that geometry is interchangeable with physics.

By accepting GR, physicists sacrificed the great scientific achievement of the 18th and 19th centuries, the laws of conservation of energy-momentum and angular momentum, which, as Emmy Noether's theorem states, are associated with the properties of spacetime symmetry: homogeneity of spacetime and isotropy of space~ \footnote{Max Planck considered the basis for conservation of momentum in Newtonian mechanics to be the action--reaction principle \cite{Planck}.
This principle may well be generalized to electrodynamics, where the role of electric charge $e$ is twofold: $e$ as a measure of the field effect on the particle ($e$ enters as a factor in the expression for the Lorentz force), and $e$ as a measure of the power of the field source ($e$ enters as a factor in the expression for the current density).
In GR, the action--reaction principle is explicitly violated.
Indeed, the motion of a particle of mass $m$ is governed by the geodesic equation $\frac{d^2z^\lambda}{d\tau^2}+\Gamma^\lambda_{~\mu\nu}\frac{dz^\mu} {d\tau}\frac{dz^\nu}{d\tau}=0$, 
which does not depend on $m$, while from the gravitational field equation with a point source $\left(R^{\mu\nu}-\frac12Rg^{\mu\nu}\right)(x)=8\pi G_{\rm N} m\int_{ -\infty}^\infty d\tau\,{\dot z}^\mu(\tau)\,{\dot z}^\nu(\tau)\,\delta^{4}[x-z(\tau)]$ it follows that the larger $m$, the stronger the gravitational field generated by this particle.
The gravitational field acts on any particle in the same way, regardless of $m$, and the influence of the particle on the gravitational field is different for different $m$.
The principle of equivalence is incompatible with the action--reaction principle.
Therefore, there is no rationale for conservation of the 4-momentum in GR.}.
Incredible efforts were made to save these laws within the framework of GR, mainly for the ``island Universe'' model, in which all matter is compactly concentrated in a finite part of space, and there is vacuum around it.
In this model, for any structure of the ``island'', the space is always asymptotically flat, which serves as a prerequisite for the Hamiltonian description of this system \cite{ADM-1}--\cite{ADM-3}.

But even having formally defined the Hamiltonian, we are not able to close our eyes to the fact that time in the Universe evolving in a non-stationary mode cannot be uniform.
First, the selected point on the time axis is the moment of the Big Bang.
Secondly, there is a permanent accumulation of matter and the subsequent collapse with the formation of black holes, and this process is irreversible.
Thirdly, the topological complexity of a system containing black holes leads to a break in timelike directions, more precisely, to the phenomenon of geodesic incompleteness.

The notion of asymptotically flat space does not have a mathematically unambiguous definition.
Qualitatively, we are talking about the disappearance of curvature at spatial infinity, $R^\alpha_{~\beta\gamma\delta}\to 0$, $r\to \infty$, and about the choice of coordinates, which would ensure the convergence of additive quantities, such as Hamiltonian and Lagrangian.

However, these requirements are not restrictive enough to eliminate the arbitrariness in foliating the space-time bundle into space-like sections consistent with the asymptotic flatness condition.
In particular, the freedom to choose a scale grid leads to the fact that the energy of a Schwazschild black hole generated by a particle of mass $m$ can take any value greater than or equal to $m$ \cite{Denisov}.

The conclusion about the presence of ill-defined additive quantities in GR resembles the Banach--Tarski theorem \cite{Banach}~\footnote{The book \cite{Wagon} is devoted to a detailed modern discussion of the Banach--Tarski theorem.}.
According to this theorem, a three-dimensional ball can be split into a finite disjoint subsets which can then be put back together through continuous movements of the pieces, without changing their shape and without running into one another, to yield a ball twice as large as the original.
The measure appearing in the Banach--Tarski theorem is the ordinary volume of the balls, more precisely,  Lebesgue measure, while the  measure in the black hole energy problem is the measure of a linear functional whose form stems from the Arnowitt--Deser--Misner Hamiltonian \cite{ADM-1}--\cite{ADM-3}.

It is not customary to discuss the Banach--Tarski paradox in a physical context for the banal reason that macroscopic bodies are made up of atoms.
The  partitioning procedure of a mathematically continuous ball has nothing to do with disintegrations of solids.
It is impossible to cut up a pea into several pieces and then reassemble them to form a  Sun-sized ball.
But this reasoning overlooks one important case,  a black hole.
Each isolated stationary black hole is completely specified by three parameters: its mass $m$, angular momentum $J$ and electric charge $e$.
Whatever the content of a system which collapses under its own gravitational field, the exterior of the resulting black hole is described by the Kerr--Newman solution.
A black hole has no granular structure, it appears as a perfect object.
Unlike ordinary quantum-mechanical bound systems, black holes are devoid of discrete energy spectra.
This suggests that the measure of the linear functional in the black hole energy problem  can be subject to transformations in the spirit of the Banach--Tarski theorem.

Gravity in GR is not a usual physical field with uniquely defined energy-momentum and evolution taking place in Minkowski spacetime ${\mathbb R}_{1,3}$.
In this, it radically differs from other three fundamental interactions in the SM.
Is it possible, following the Poincar\'e tenet mentioned above, to return to the field-theoretical treatment of gravity, choosing ${\mathbb R}_{1,3}$ as the arena of all physical processes?
Such treatments are indeed feasible \cite{Rosen} if the gravitational field, described by the second rank tensor $\phi_{\mu\nu}$, can be granted to be always ``sufficiently weak'', {\it i. e.}, in the relationship $g_{\mu \nu}=\eta_{\mu\nu}+\phi_{\mu\nu} $, where $\eta_{\mu\nu}$ is the metric tensor of the flat background ${\mathbb R}_{ 1,3}$, the field $\phi_{\mu\nu}$  has small components in every inertial frame of reference, $|\phi_{\mu\nu}|\ll 1$.
The homogeneity of ${\mathbb R}_{1,3}$ affords the energy-momentum conservation by virtue of the standard arguments of Noether's first theorem.
The field-theoretic treatment  remains valid as long as the mapping $g_{\mu\nu}\mapsto\phi_{\mu\nu}$ is a bijective smooth mapping, in other words, any allowable configuration of a manifold with the metric $g_{\mu\nu }$, associated with the gravitational effect, can be smoothly covered by a single coordinate patch.
However, solutions of the dynamic equations in GR are given by manifolds pitted by black holes.
Attempts to come up with a mechanism that excludes solutions with nontrivial topology \cite{Logunov} led to field theories which did not find wide acceptance.
Should such solutions be excluded altogether?
A reasonable assumption is that a black hole in the field-theoretic framework is to be  regarded as a singular solution, like a spherical shock wave.
It is well known that the presence of shock waves in gas dynamics is quite compatible with the laws of conservation of energy and momentum.

\subsection{Those physicists--vermins have imparted wrong spin to the ball on a bet~\cite{Galich} }
\label
{This vermins bet untwist balls}
The world population today has 8 billion inhabitants.
A necessary condition for its prosperity is the power engineering development.
The world is sophisticated enough in both the laws of physics and the Earth sciences to realize a simple truth: we produce and consume energy in a highly unwise way.
It is time to finally stop burning fossil hydrocarbons: coal, oil, and gas.
Network of nuclear power plants~\footnote{This subsection uses conventional abbreviations: NPP, nuclear power plant, MOX fuel, fuel containing oxides of fissile materials ({\bf M}ixed-{\bf Ox}ide fuel), SNF, spent nuclear fuel.} working in a closed fuel cycle is the optimal solution for the entire Earth for the coming millennia.
With regard to the neutron spectrum, nuclear reactors are divided into two types: thermal neutron reactors (or thermal reactors) and fast neutron reactors (or fast reactors). 
Almost all power reactors (commercial and military) made to date are thermal.
The fuel for them is the uranium isotope ${}^{235}{\rm U}$.

Natural uranium is 99.3{\%} of ${}^{238}{\rm U}$, while ${}^{235}{\rm U}$ is only 0.7{\%}.
But the fuel must contain at least 3--5{\%} of ${}^{235}{\rm U}$.
Fuel of this composition is obtained from natural uranium by ``enriching'' it.
A 1 GW thermal reactor is loaded with 20 tons of enriched uranium.
About a year later, the fuel is replaced with fresh one.
During the year, only 1 ton out of 20 burned out, but it still needs to be replaced, since the reactivity of the fuel has changed.
It is difficult to further work with it.
There were 19 tons of radioactive SNF.
We should also take into account that the initial 20 tons of enriched uranium were obtained from 200 tons of natural uranium, in other words, 180 tons of uranium after the enrichment procedure go to the dump.
So, by burning 1 ton of uranium, we produced 199 tons of wastes: 180 tons of the enrichment dump and 19 tons of SNF.

The terrestrial supply of ${}^{235}{\rm U}$ is rather small.
If we proceed from explored deposits, then its reserves at the current rate of consumption of this isotope run out before oil and gas reserves do.
SNF from thermal reactors contains re-usable plutonium (the share of ${}^{239}{\rm Pu}$ in it is 60{\%}, the rest refers to higher isotopes of Pu).
Plutonium can also burned in thermal reactors, but here it is less effective than ${}^{235}{\rm U}$, and the process is more expensive.
Processing and storage of thousands of tons of SNF, liquid waste from its processing and accumulated plutonium, as well as mountains in enrichment dumps is one of the most costly parts of nuclear energy.
Storing 50 tons of plutonium costs $\sim \$100$ million a year.
SNF is usually stored at the same nuclear power plant where it was produced.
This state of affairs cannot last forever.
Therefore, we are looking for opportunities for its geological burial.
However, finding a burial place and obtaining the consent of the local population for the construction of a nuclear waste repository is not an easy task.
As a rule, the decision is simply postponed for the future.

The fuel for fast reactors is a mixture of plutonium and ${}^{238}{\rm U}$.
Such reactors can produce from ${}^{238}{\rm U}$ as much plutonium as they consume.
The reactors need an initial portion of plutonium, and then they only need to add ${}^{238}{\rm U}$ which is cheap and has a large supply.
A regime is possible that both electricity and an additional portion of plutonium are produced -- for other reactors or weapons.
The reactor of this kind is called a breeder.
A mode is also possible when the amount of plutonium decreases; such a system is called a burner.
In a word, the problem of depletion of ${}^{235}{\rm U}$ reserves does not exist for fast neutron reactors.
They actually work using ${}^{238}{\rm U}$.

A reactor has also been developed in which the consumable fuel is a widely distributed in nature thorium isotope ${}^{232}{\rm Th}$.
This reactor can operate on both fast and thermal neutrons.
The explored reserves of ${}^{238}{\rm U}$ and ${}^{232}{\rm Th}$ are enough for the entire world energy industry during many millennia.

Fast-neutron reactor technology answers the question of what to do with the plutonium accumulated in spent fuel from thermal reactors and huge amount of ${}^{238}{\rm U}$ in enrichment dumps.
Everything can be used as fuel for fast reactors.
Fast reactors mitigate one of the most acute troubles of nuclear energy, the problem of accumulation of SNF.
When plutonium and minor actinides are extracted from SNF, which will be reloaded into the reactor as fuel, the radioactivity of SNF, consisting of only fission fragments, is reduced tenfold.
The issues of waste disposal without environmentally hazardous consequences are now being solved.

A fast neutron reactor is more complicated than a thermal neutron reactor.
It cannot be cooled with water: water slows down neutrons well.
A coolant made of relatively heavy elements is called for.
The researchers tried mercury, a mixture of sodium and potassium, a mixture of bismuth and lead, until they settled on sodium.
Although sodium is flammable and actively reacts with water, when having regard to the combination of technological, economic, and safety factors, this coolant is by far the most optimal.
The lead-cooled reactor seems attractive due to its simple design.
In Russia, power reactors with sodium coolant are operated: since 1980, BN-600 for 600 MW, and since 2016, BN-800 for 880 MW.
A lead-cooled reactor, BREST-300 for 300 MW, is being built~\footnote{A number of experts are skeptical about its construction.
They believe that new elements, the nitride fuel ${\rm (U_{0.8}Pu_{0.2})N}$ and lead coolant, should be studied first in small reactors.
To be specific, the poorly studied problem of thermochemical instability of nitrides (threatening swelling of the fuel core and corrosion of the cladding), and that of instability of the oxygen potential (threatening uncontrolled deposition of lead oxides on the walls of a double-loop cooling system) are alarming.
Radiochemical processing has not been tested on miniature prototypes of this system.}.
It is planned to construct BN-1200 for 1200 MW.

One of the problems of fast neutron reactors is the difficulty of preparing fuel for such reactors.
Another serious problem is the ability to burn out the $\alpha$-active ${}^{241}\!{\rm Am}$ in current fast neutron reactors.
Closing the nuclear fuel cycle requires a high rate of ${}^{241}\!{\rm Am}$ burnout by the reactor and a short duration of the radiochemical SNF reprocessing cycle.
At the BN-600 and BN-800 reactors, this goal is not achieved~\footnote{At the current rate of radiochemistry, to be precise.} because the rate of burning ${}^{241}\!{\rm Am}$ here does not exceed the rate of its formation from ${}^{241} {\rm Pu}$.
Therefore, additional burners are essential.
Theoretically, BREST-300 is capable of burning out not only the ${}^{241}\!{\rm Am}$ formed in it, but also up to $30 \%$ of americium loaded into the reactor from outside.
Whether this forecast will be justified in practice, we will find out in the near future.

As for ${}^{232}{\rm Th}$, it does not form rich deposits, unlike uranium, although the reserves of thorium in the Earth's crust are 3--4 times higher than those of uranium.
The technology for extracting ${}^{232}{\rm Th}$ from ore is rather complicated.
Currently, a single 40 MW reactor with fuel containing ${\rm ThO_2}$ is operating in India.
It is not normal.
Thorium reactors need to be dealt with in full force.
  
The breeder reactor makes it easy to accumulate ${}^{239}{\rm Pu}$ in amounts needed to make a nuclear weapon.
To prevent the emergence of new countries (or large terrorist groups) possessing such weapons, the United States tried to impose an international ban on any activity in the development of fast reactors, including research-type systems.
All Western countries, except France, obeyed the ban.
In the 1960s and 1970s, France successfully coped with the problem of making MOX from ${\rm (U_{0.8}Pu_{0.2})O_2}$ and put into commercial operation 
sodium-cooled fast neutron reactors: in 1974, {\it Phoenix} at 250 MW, and in 1985, {\it Superphoenix} at 1200 MW.
However, due to technical blunders, administrative and political reasons, the reactors were stopped, and by the end of the century, their activities were terminated.
At present, only Russia has fast-neutron power reactors in operation.

Nuclear power originated as a by-product of the military program.
When the military technology of thermal neutron reactors reached the industrial level, the idea arose to use these achievements in the energy sector.
At that time, there were two types of reactors: gas-graphite for the production of plutonium and pressurized water (in which water takes heat from the fuel and slows down neutrons) for submarines.
Reactors of the first type were not very reliable for power engineering, and water-cooled reactors proved to be better.
On their basis, the industry of nuclear energy was founded.

For a quarter of a century, this line of development has been rapidly developing in the USA, the USSR, and Western Europe.
It malfunctions as a result of the accident at the American nuclear power plant Three Mile Island in 1979.
Work to eliminate its consequences lasted 14 years and required $\sim 10^9$ dollars.
Safety requirements have risen sharply, the cost of building a nuclear power plant in the United States has jumped by an order of magnitude.
Before the accident, 400 new power units were laid down in addition to the 100 already in operation; after the accident, all these projects were stopped as unviable.
Since then, the number of power units in the United States has remained almost unchanged, mainly due to the decision about the extension of the forty-year period of their operation for another 20 years, but the share of nuclear power plants in the energy sector is constantly decreasing.

In 1986, there was an accident at the Chernobyl nuclear power plant, and in 2011, an accident at the Fukushima-1 nuclear power plant.

The world learns the consequences of accidents and assesses its own willingness to pay for the entire life cycle of such systems, taking into account the costs of additional safety systems and the problem of SNF accumulation.
Developed countries refuse to commission new power units, some fix the refusal by law.
The American concern Westinghouse was sold to the Japanese company Toshiba; the British corporation BNFL was attempted to be privatized, but there was no buyer.
France divides the state company Areva into parts, transferring its management to other companies; Germany's Siemens is curtailing its nuclear business.
The era of power engineering with thermal neutron reactors is nearing its logical conclusion

The development of nuclear energy, which is based on thermal neutron reactors, is a historical mistake.
The cost of this mistake is skyrocketing.
The sooner humanity will completely switch to energy using thorium reactors and uranium fast neutron reactors with a closed cycle, the sooner it will acquire the features of a single intelligent organism.

It is obvious that ``green'' energy (wind turbines, hydroelectric power plants, solar panels), no matter how successfully it develops, will be able to satisfy only a small fraction of the energy demand, only in some places on Earth (with climate change, the geography of these places will be unpredictable shift).
On the other hand, to continue to burn coal, oil, and natural gas is to shamelessly rob our descendants.
We need to stop this process on an international scale.

A serious challenge for applied physics and technology in the near future will be the search for ways of transmitting electrical energy over long distances without significant ohmic losses.
It is interesting that, parting with ``bomb-making'', Sakharov pondered over the prospect to pursue this matter on a full-time basis, but circumstances pushed him to human rights and political activities.

On the way to total nuclear energy, we may face surprises and troubles.
The problem of closing the nuclear fuel cycle is very complex.
At every step there are many unknown paths; the correct choice of one of them requires not only the presence of intelligence, experience, and will, but also patience (trials last for years), coherence of efforts, and a systematic strategy.
The solution of this problem in the US began in the late 1940s; the USSR were seriously engaged in this for several decades by constructing the BR-2, BR-5, BR-10, BOR-60, BN-350, and BN-600 systems.
With a responsible approach to the energy of the future, we must solve the problem of closing the nuclear fuel cycle.
Another way of looking at it is nonexistent.

An integral attribute of modernity has become automobile, air, and water transport, driven by the burning of petroleum products.
What will replace it in the transition to a purely electric way of life?
Various alternatives can be discussed, such as the use of hydrogen produced by decomposition of water, or liquid fuels of agricultural origin, but, apparently, it will be more appropriate to notice two general indisputable facts here.

First, there is a need in cities, especially megacities, to move almost all masses of people by public transport on electric actuation mechanism.
Secondly, the development of robotics and communications makes the physical presence of people in their places of work less and less necessary.
Their movement over long distances by high-speed trains can in most cases be preferable to air flights from the point of view of efficiency and economy.

On the whole, this apparently threatens to infringe on the customary individual rights of a {\it chosen part} of the current bourgeois world in favor of another part which is actually deprived of most of the means of civilized subsistence.

\section{Peculiarities of the national  science}
\label
{Features}
It is widely believed that science is international.
One cannot but agree with this if one bears in mind its international mission.
The twentieth century demolished many barriers to scientific research, and representatives of different races and states joined it.
However, now the ideals of internationalism in the functioning of science are being seriously tested due to the conflict between Russia and the West.
The scientific community may not be able to cope with this test.
A recurrence of national strife loomed.
In order to counter this, one should at least not hush up the sources of this disease, obvious and hidden.

The manifestation of national character is indelible in any sphere of human activity.
Physics is no exception.
Consider, for instance, the problem of confinement.
The very name of this phenomenon and the ``coercive, forced'' approach to its interpretation \cite{Wilson74} reveals a characteristic yankee style.
The Japanese were not satisfied with this treatment.
Overcoming the feeling of shame occupies a major place in their spiritual life.
The color nature of quarks and gluons is ``shameful''.
They need a means of its reliable concealment.
And the means ($Q_B|{\rm phys}\rangle=Q_C|{\rm phys}\rangle=0$) were found \cite{KugoOjima}.

But jokes aside.
Discussing ethnic themes judiciously is difficult and dangerous.
It is, however,  urgently needed now.

The division of science into ``Aryan'' and ``non-Aryan'' in Nazi Germany was not out of the ordinary.
Anyone who thinks that this cannot happen in countries that have reached the modern level of civilization, and certainly in the Soviet Union, a stronghold of friendship between peoples, is mistaken.
In the mid-1930s, Landau could have said to students at the Kharkov Mechanical 
Engineering Institute: ``Theoretical physics is not an occupation for the Slavs''~\cite{Josephson}.
Of course, this arrogant passage could be attributed to a one-sided development of the child prodigy and his little life experience.
After 20--30 years, it would not have occurred to him, and not even because he spent a year in prison, which taught him restraint in front of strangers.
He met colleagues of Slavic origin (Bogoliubov, Sakharov), whose achievements in theoretical physics were not only not inferior in depth and significance, but in a number of cases far exceeded his own.
A friend of his youth, Georgi{\u\i} Antonovich Gamow, whom he seemed to know inside out, eventually appeared, as a scientist, to be more far-sighted and more productive than Landau.

Young Landau with his bad manners is not a single example of a skeptical and slighting attitude towards Slavic talent.
This is a system of views rooted both in the West and in our country.
The image of its Russian bearer, Smerdyakov, is captured in the novel {\it The Brothers Karamazov} by Fyodor Mikhailovich Dostoyevski{\u\i}.

Mind itself is not among the main qualities brought up in a Russian child.
``Had three sons that father, the eldest was clever one; so-so, was middle their brother, and fool was the youngest son!'' \cite{Ershov}.
Indeed, neither in the folk epic nor in Russian literature is there a hero who, like Homer's Odysseus, would not only achieve success in life, but also decide the outcome of the confrontation of peoples due to exceptional mental resourcefulness.

Among the Slavs, the meaning of the mind does not lie on the surface.
They value intelligence only when paired with honesty, courage, and commitment to good deeds.
Outside of this combination, other terms are used for manifestations of the mind.
The Hellenes were less exacting.
They were not embarrassed by the opportunism of Odysseus.
Let us recall at least how he tried to slope off from the campaign against Troy.
Another imperative is not to waste your mind on trifles ({\it aquila non captat muscas}~\footnote{Eagle does not catch flies.}).
A talent meant for big things~\footnote{Had Homer lived these days, he would not have missed the opportunity to send Odysseus to Russia.
This is his final destination.
The hero shakes off the remnants of the playboy attributes; here he cognizes life in its tragic grandeur.
Odysseus takes a liking to Russia with his heart: ``Farewell to Ithaca! 
Farewell to dear Penelope!''
The Russian incarnation of Odysseus would not even need to be invented.
Grigori{\u\i} Yakovlevich Perelman would be quite suitable as a prototype.}.
And if you managed to solve them, then to bother about honors means to humiliate your talent.

In the West, the features of the Russian worldview are often either not understood or ignored.
A classic example is the attitude to the discovery by Dmitri{\u\i} Ivanovich Mendeleev of the periodic law of chemical elements, on the basis of which he predicted in 1869 the existence and properties of then unknown elements: gallium, germanium, scandium, and calculated their atomic weights.
The discovery of this fundamental law of nature was not awarded the Nobel Prize \footnote{Mendeleev died in 1907. 
The Nobel Prizes in Chemistry in 1901--1906 were awarded for results of a much smaller scale.}.
And even the Mendeleev periodic table is called {``the periodic table''}, without mentioning the name of its author.

The history of science knows many examples of a biased attitude towards achievements of the Russians.
Someone may object: ``Is it worth turning over the pages of ancient times with their rude customs?''
OK.
Let's go back to the civilized period of history.

The physical mainstream today is determined by astrophysics and cosmology.
But nowhere will there be even a hint to call the founding fathers of this mainstream people from Russia.
Often, their pioneering findings caused distrust or ridicule.

In 1922 Aleksandr Aleksandrovich Friedmann published his theory of non-stationary Universe \cite{Friedmann}.
Einstein called it suspicious \cite{Einstein1}, but later admitted he was wrong \cite{Einstein2}, and declared the cosmological constant introduced by him to be his own greatest scientific blunder.
Sakharov showed that if we take quantum theory into account, the presence of this constant is unavoidable~\footnote{The qualitative explanation is as follows.
The positive energy of zero-point oscillations of bosonic fields and the negative energy of the Dirac sea of fermion fields taken together give rise to a Lagrangian containing the scalar curvature of spacetime and the cosmological constant, in other words, the vacuum of matter induces the elasticity of spacetime, which manifests itself as gravity.}~\cite{Sakharov}.
So, Friedmann established what is the realistic mode of {\it evolution} of the Universe.
His disciple Gamow supplemented the description with thermodynamics, introducing the concepts of {\it Big Bang} and {\it Hot Universe}.
He calculated the proportions of light elements that appeared in the primordial nucleosynthesis \cite{Gamov46}--\cite{Gamov49}, predicted the {\it cosmic microwave background}, and in 1953, estimated its {\it temperature} at 3 ${\rm K}$ \cite{gamow53}, \cite{Chernin}.
These discoveries were put by his American colleagues on a par with jokes which gushed from Odessite Gamow.
The ideas of the Big Bang and the Hot Universe have now received universal recognition, and measurements for the temperature of the cosmic microwave background gave 2.7 ${\rm K}$.
In 1967, Sakharov found an explanation why the early Universe, containing matter and antimatter in equal proportions, lost this symmetry, appearing to us filled only with matter \cite{Sakharov67}.
In the cosmology of the early Universe, the mechanism of {\it  baryon asymmetry} became a cornerstone.
In 1965, Erast Borisovich Gliner suggested that the cosmological constant should be regarded as the quantity responsible for {\it antigravity} in the model of an empty {\it inflating Universe} with the de Sitter metric and the equation of state $p=-\varepsilon$, where $p$ is pressure and $\varepsilon$ is is the vacuum 
energy density.
With this, he came to Zel'dovich at the Institute of Applied Mathematics.
Igor' Dmitrievich Novikov, who worked as an assistant to Zel'dovich, stood in his way.
Novikov did not like Gliner's ideas because they contradicted the canons of theoretical physics outlined in the course of Landau and Lifshitz.
Gliner (who lost his arm on the fronts of the Great Patriotic War and spent 10 years in work camps) was not embarrassed.
He removed Novikov and appeared before Zel'dovich who did not appeal to Landau's authority but only remarked to Gliner that the imaginary value of the speed of sound follows from his equation of state.
On that they parted.
Stephen Hawking visited Moscow.
In a conversation with him about the creation of particles in a gravitational field, Gliner's equation of state surfaced, and Zel'dovich repeated his criticism to Hawking.
But Hawking replied that a vacuum need not be viewed as a medium through which sound waves travel.
Zel'dovich agreed.
Gliner's paper \cite{Gliner} was published.
Now the cosmological constant is considered the embodiment of antigravity.
Its nature is associated with {\it dark energy}.
And the inflation of the early Universe was rediscovered 16 years later by theorists of the USA and the USSR.
Now it is a settled element of cosmology.

Do these grandiose discoveries bear the names of their authors?
The only one is Friedmann's result.
But a tail of other names is attached to it.
At the risk of breaking the rules of good manners, calling the standard cosmological model the {``world of Friedmann, Lemaitre, Robertson, and Walker''} is, in my opinion, about the same as pronouncing the name of a Napoleonic marshal separated by a comma from the names of sutlers.

So far we have been talking about the theoretical foundations of cosmology.
But still experimental discoveries in astrophysics and cosmology belong to the West.
May it be that the Russians do not know how to combine the work of the brain with the work of the hands?
Obviously, such is not the case if we talk about things serious.
During the Great Patriotic War, the Red Army had weapons that were not inferior in quality to the weapons of the Nazi army, and Wehrmacht generals could only dream of rocket artillery (``Katyusha'').
With terrible human losses and a destroyed post-war economy, the USSR managed to develop a nuclear industry and rocketry on the same level as the United States.
It took ingenuity and coherence of large teams.
Everything had to be done from scratch by themselves; there was simply no one to buy or borrow from.
I knew these skillful people, I saw the wonders of their skill.
In the mid-60s, North Vietnam fought the United States with Soviet weapons, and won.

With regard to experimental discoveries in fundamental research, here our lag is due to the scarcity of resources.
I would like to refer to the director of VNIIEF, according to whom the budget of the institute in the late 70s was approximately equal to the budget of the USSR Academy of Sciences with its 330 scientific institutions and a staff 10 times larger than the staff of VNIIEF.
In fact, the experimental lag in fundamental areas is not an entirely simple matter.
Let us take just one example.

Mikhail Evgen'evich Gertsenshtein and Vladislav Ivanovich Pustovoit proposed to use a laser interferometer for detecting  gravitational waves in their 1962 paper \cite{GertsenshteinPustovoit}.
Ten years later, a modern textbook {\it Gravitation} issued \cite{Misner}, in which chapter 37 is devoted to gravitational waves and different ways for their detection.
There is not a word there about the Gertsenshtein--Pustovoit method.
However, it is this method which was used by  the Laser Interferometer Gravitational-Wave Observatory (LIGO) Collaboration to detect gravitational waves in 2015.
The 2017 Nobel Prize for this discovery was awarded to three American members of the LIGO Collaboration: Barry Barish, Rainer Weiss, and Kip Thorne.
Although Pustovoit was alive at this time, the Nobel Prize Committee did not remember him.
And how did the prizewinners speak of the source of their triumph in their Nobel Lectures \cite{ThorneNobel}? 
Barish did not mention it at all.
It may be concluded from his Lecture that Barish is a boss responsible for financial and administrative affair.
Weiss restricted himself to the remark that, early in the 1970s, he was pondering over the possibility to arrange the Michelson interferometer for detecting gravitational waves, and unknown to him two Russians,  Gertsenshtein and Pustovoit, had come up with a similar idea in 1963~\footnote{Actually, in 1962.
English translation of the JETP issue with the paper \cite{GertsenshteinPustovoit} indeed appeared in 1963, but the date of the original publication of this paper, 1962, was indicated in it.}.
Thorne claimed that he regards Weiss as the primary inventor of gravitational interferometers because  Weiss identified the most serious noise sources, described ways to deal with each one, and proposed kilometer-scale arm lengths of an L-shaped laser interferometer design~\footnote{Why did Gertsenshtein and Pustovoit consider a device of size of $\sim 10$ m? 
A major reason is that optical lasers were  first constructed two years prior to publication of their paper.
It is then clear what reasoning about the laser power was advisable at that time.
Making kilometer-scale arm lengths of a device requires regularly operating lasers with $\sim 200$ W radiation power.
A laser of this kind became available much later. 
It was invoked as a key element of the LIGO project.
As to the noise sources, one was to traverse a long path toward their identification and overcoming in which a central role again played the Russians, Vladimir Borisovich Braginski{\u\i} and his colleagues.
This fact is openly stated in \cite{Thorne}.} in an internal MIT report \cite{Weiss}.
According to Thorne, Gertsenshtein and Pustovoit proposed the bare-bones idea of a device earlier and independently of  Weiss.
One cannot argue with that.
Indeed, Einstein advanced the bare-bones idea of gravitational waves  earlier and {independently} of 
Thorne, and Thorne regards himself as an icon for the LIGO Collaboration \cite{ThorneNobel}, 
which was the first to observe gravitational waves~\footnote{Einstein thought of such waves as too weak entities to  be ever detected.}.

\section{Tyapin Island}
\label
{Tyapin}
Owing to the discovery of nuclei with atomic numbers $114$ and $116$, made in Dubna by the international teams of experimenters led by Yuri{\u\i} Tsolakovich Oganesyan~\cite{Oganessian}, \cite{Yakushev}, we learned that the Mendeleev table of elements does not end at $Z=105$ or so.
A little further away, the outline of an island of relatively long-lived nuclei looms.
The existence of this island was not a surprise.
It was predicted a long time ago,  but it was possible to confirm the prediction only at the beginning of the 21st century.
Who was the predictor of the island?
The Internet hints at Georgi{\u\i} Nikolaevich Fl\"erov.
Most likely, Fl\"erov really spoke out on this score, but for him, as an experimenter, the matter could be limited to intuitive guesses related to many years of experience in research into the synthesis of 
transuranium nuclei.
As to a clear theoretical prediction, its author is Anatoli{\u\i} Sergeevich Tyapin.
His article~\cite{Tyapin} states: ``In the range $Z=114-124$, there is no pronounced magic number, but all nuclei with the indicated charge values will have relatively greater stability than 
neighboring nuclei with other charge values''.
Although the prediction of the group of relatively stable heavy nuclei is said to have been in the 
air \cite{Malik}, in 1971, this was the only rigorous result obtained using the whole 
mathematical power of the shell correction method developed in pioneer 
works~\cite{Tyapin-1964}--\cite{Tyapin-1974}, \cite{Tyapin}.

Researchers immersed in the topic are familiar with this prediction and its theoretical basis, but avoid explicitly mentioning it.
How to understand this?
The objective reason is that the discovery was ahead of its time, because its experimental verification becomes possible after three decades.
Probably, for this reason,  in 1971, the nuclear community did not attach importance to this discovery, and later on, they forgot it; fortunately, the author did not blow about it.

This is reminiscent of the story of how Oliver Heaviside introduced the delta-function (under the name {impulsive function}) together with its Fourier transform in 1899, laying the foundation of operational  calculus.
However, this discovery exceeded the requirements of physics at that time, and went unnoticed.
Thirty years later, Paul Dirac came up with the same notion.
Today a physicist will say: ``the Dirac delta-function'', and he couldn't care less.
Let it be!
At least the antiderivative of $\delta(x)$ has got the name of the Heaviside step function \cite{Before}.

There is also a subjective side of the matter.
I have known Tyapin for many years, and I would like to share my thoughts.
Echoing Pushkin's ``the devil guessed I was born in Russia with soul and talent'', one can say about Tyapin: ``The devil guessed he was born terribly quick-witted and capable of scanning human souls''.
Tolya was a child prodigy, but he was not born into a Jewish family ready to raise a child prodigy~\footnote{It is likely that, in ancient times, close attention to the mental characteristics of a newborn among the Jews was associated with the expectation of the advent of the messiah in human form, but in our time, it has mostly lost its religious meaning and preserved as a household tradition.
Jewish surnames such as Liebersohn (Beloved Son) and S{\"u}sskind (Sweet Child) serve as reminder of this custom.} and patiently treat a disease typical of such creatures, conventionally designated by the term ``arrogance'', because an unhealed child prodigy can remain an infantile, bringing cruel blows of fate on his head.
You don't have to go far for examples.
The 24-year-old Landau taught the highly respected patriarch of Soviet physics Abram Fedorovich Ioffe: ``Theoretical physics is a complex science, not everyone can understand it'' (the maxim is essentially true, but uttered in an obscene tone), for which he paid with his position of an employee in the Leningrad Physics \& Technology Institute \cite{Bessarab}.

About three decades later, a similar story happened to Tyapin when he was studying at MEPhI.
His supervisor Alexander Solomonovich Kompaneets~\footnote{Landau's disciple, the first to pass all exams according to the famous ``theor-minimum''.} offered Tolya tasks of ever-increasing complexity and was astonished how quickly he coped with them.
Finally, he gave Tolya a problem in plasma physics that Landau himself could not solve.
Tolya instantly discovered a stumbling block and outlined a way to overcome the difficulty.
To celebrate, Kompaneets dragged him to Landau.
Tolya, with the tactlessness characteristic of the young talent, showed the master where he stumbled and how he should have acted in order to succeed.
Landau silently swallowed this challenge and said that he did not want to see Kompaneets' prot{\'e}g{\'e} in the future.
With a heavy heart, Alexander Solomonovich realized that, from now on, in the Moscow academic environment, to open his pet's talent will be impossible.
All he could do was to persuade Tolya to take assignment to work in the closed Arzamas-16, so that there, from an experienced teacher Baz', he could learn a professional skill in working in nuclear theory.

Baz' was a benevolent person with a subtle sense of humor \cite{BazUFN}.
However, Tyapin revealed immediately in him something deeply hidden from most of those around him, and this gave rise to mutual repulsion.
Baz' avoided conflicts.
He invited Tyapin to go to the Kurchatov Institute for a one-year internship, to find there both theoretical work to your liking and comrades to your liking.
Tyapin agreed.

At the Kurchatov Institute, fate brings Tyapin to Vilen Mitrofanovich Strutinski{\u\i}, while who tried to combine the semi-empirical drop model of the nucleus with the more fundamental quantum mechanical shell model.
They successfully overcome difficulties within a month, and publish the article \cite{Tyapin-1964}.
However, their further cooperation does not add up.
Strutinsky had a reputation as a pleasant person in every respect \cite{Brack}, but even on this ``Sun'', Tyapin managed to discern spots.
Tyapin interrupts his internship and returns to VNIIEF.
He already felt like an independent researcher of the nucleus.

The 1960s were a period of farewell to nuclear theory as a physical mainstream.
Most of the recognized leaders in this field switched to research in the theory of elementary particles, astrophysics, and cosmology.
Tyapin could also make a turn into one of these lanes.
But he followed his own course, always at the forefront of nuclear research which acquired mathematical rigor and physical solidity in his articles.
Of course, a lot of effort and time was spent on calculations related to the plans for closed work of VNIIEF.

The range of his interests was quite wide.
He studied the propagation of seismic waves in a two-layer medium, developed methods of numerical calculations for kinetic models of destruction of materials.
Nevertheless, his scientific passion was nuclear theory.

Tyapin understood that no fundamental description of the nucleus had been made.
There is only a fragmented collection of models adapted to explain individual sets of phenomena.
QCD confidently defended its claims to the status of an officially recognized theory of strong interactions at high energies $E\gg 100$ MeV.
But the Yukawa mechanism, according to which nucleons are kept in the nucleus due to the exchange of mesons, has so far remained the pillar of nuclear physics~\footnote{The Yukawa mechanism, equipped with a number of innovations, such as spontaneously broken chiral symmetry, effective Lagrangians, and derivative expansions~\cite{Weinberg1990}, dominates this area to the present day~\cite{Epelbaum}, \cite{Machleidt}, although the issue of understanding nuclei in terms of quarks was high on the agenda of the QCD developments since the 1970s.
The simplest possibility was to think of a nucleus with mass number ${\cal A}$ as a bound system of ${\cal N}=3{\cal A}$ quarks enclosed in a ``bag'' of size $R$ ~\cite{Petry}.
But the stability of the bag stipulates that ${\cal N}$ and $R$ must be related by $R\sim{\cal N}^{\frac14}$, contrary to the firmly established phenomenological relation $ R\sim {\cal N}^{\frac13}$, and this discord is particularly sticking for heavy nuclei.
Furthermore, the magnetic moments of such bags differ from the experimentally measured magnetic moments \cite{Arima}, \cite{Talmi}.
An effort to account for the static properties of nuclei by eliminating gluon degrees of freedom was reasonably successful~\cite{Maltman}, but never progressed beyond small nuclei, D, T, and He.}.
In the late 1980s, when the entire arsenal of the Yukawa concept was essentially exhausted, Tyapin came to the conclusion that further advances in nuclear physics are impossible without a radical update of its foundations, specifically by invoking an effective theory to low-energy QCD.

Unfortunately, at this time he has serious health problems.
Tyapin's last decade closely resembles  Landau's last six years in the sense that none of them could put into action all the richness of mind and will inherent in them from birth.
However,  there were some differences.
In 1962, Landau was awarded the Nobel Prize.
In 2012, I become aware of the existence of the Lise Meitner Prize, and I am trying to nominate Tyapin as a scientist who theoretically predicted an ``island of stability''.
Conditions of the competition included providing feedback from three well-known specialists in nuclear theory.
At that time, there were very few suitable people left.
In addition, I learned about this award by accident and belatedly.
I just didn't have enough time before the application deadline to contact three authoritative nuclear scientists and organize the writing of reviews.
A little later that year, Anatoli{\u\i} Sergeevich died.
\begin{figure}[htb]
\centerline{\includegraphics[height=0.3\textwidth,angle=90]{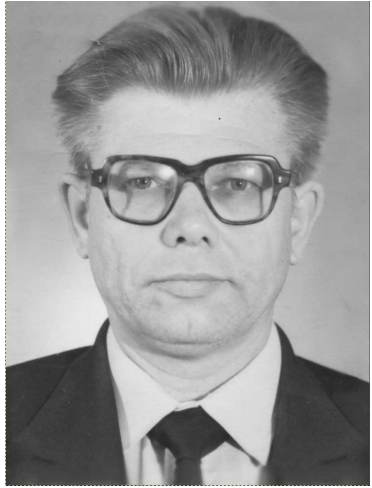}}
\caption{Anatoli{\u\i} Sergeevich Tyapin (23.01.1938 - 12.07.2012)}
\label{TyapinAS}
\end{figure}

Tyapin, as a person, was a bunch of contradictions.
With a Slavic face,  as plain as the countryside, and with an amazing gift to see through a person; with childish straightforwardness in communication and a huge margin of tolerance for other people's weaknesses; with the individual style of the researcher and the willingness to harness himself to the work of the team.
To illustrate, I will mention only one fact: Tyapin joined the CPSU in 1979, that is, a year before the proclaimed complete implementation of communism in the USSR, the failure of which was obvious to everyone, and this step did not pursue either career or selfish goals, because Tyapin never been a boss.
He could save a colleague drowning in scientific difficulties, considering this in the order of things~\footnote{Such was the inner world of a man who met the reality of war at the age of four and grew up in the conditions of trials that befell ordinary Soviet people at that time.}.
He had no enemies.
But some perceived his very presence as a mute reproach of conscience, as a reminder of something unseemly or even shameful in their biography.
Sometimes it depended on these people whether to mark the discovery of a scientist with their mention, drawing additional attention to it, or to pass it over in silence.

The contribution of Anatoli{\u \i} Sergeevich Tyapin to nuclear physics, I think, is greatly underestimated.
It would be correct to honor this outstanding figure from the end of the ``golden age'' of nuclear physics to give some of the elements of the group of relatively stable nuclei with $114\le Z \le 124$, predicted by him, his name.

Or maybe just call the whole group {\it Tyapin Island}?

\section{Finale
}
\label
{conclusion}
I don't know how to finish the story whose happy end would be trying to make fun of reality.
It may be appropriate to turn to George Byron's famous elegy: 
\begin{quote}
{    I only know -- we loved in vain --\\ 
     I only feel -- Farewell! -- Farewell!}
\end{quote}
or Guillaume Apollinaire's farewell plaintive note: 
\begin{quote} 
{J'ai cueilli ce brin de bruy\`ere\\ 
L'automne est morte souviens-t'en\\
Nous ne nous verrons plus sur terre\\
Odeur du temps brin de bruy\`ere\\     
Et souviens-toi que je t'attends}~\footnote{
\begin{quote}
{I've gathered this spring heather \\
Autumn is dead you will remember\\ 
On earth we'll see no more on each other\\
Fragrance of time spring of heather \\     
Remember I want for you forever}
\end{quote}
Translated by A. S. Kline.}
\end{quote}

I have described a number of episodes of scientific and social life that ultimately led to the decline of High Energy Physics.
I could not pass by the participants in these events.
Whether they had a beneficial or pernicious influence on the course of events, the future will show.
I am not a moralist, I am a physicist.
For me, facts are important, their causal relationship, and the logic of their appearance.
The reader will make his own assessments.


\begin{thebibliography}{100}
\bibitem
{TOP500}
TOP500 is a program for publishing (twice a year) a list of the 500 most powerful 
supercomputers built in the world.
\end{thebibliography}
\end{document}